\documentclass[prd,aps,10pt,notitlepage,showpacs,nofootinbib,tightenlines]{revtex4}
\usepackage{mathrsfs}
\usepackage{amsmath}
\usepackage{amssymb}
\usepackage{epsfig}
\usepackage{graphicx}
\usepackage{booktabs}
\usepackage{multirow}
\usepackage{subfigure}
\usepackage{bm}
\usepackage{times}
\usepackage{braket}
\usepackage{color}
\usepackage{slashed}
\usepackage{hyperref}
\usepackage{threeparttable}
\newcommand{\beq}{\begin{eqnarray}}
\newcommand{\eeq}{\end{eqnarray}}
\newcommand{\non}{\nonumber\\ }

\definecolor{Red}{rgb}{1.,0.,0.}

\definecolor{Blue}{rgb}{0.,0.,1.}

\definecolor{nicered}{rgb}{0.7,0.1,0.1}
\definecolor{nicegreen}{rgb}{0.1,0.5,0.1}
\hypersetup{colorlinks,citecolor=nicegreen,linkcolor=nicered}
\def \cpc{ Chin. Phys. C  }

\def \epjc{ Eur. Phys. J. C }

\def \jpg{  J. Phys. G }
\def \npb{  Nucl. Phys. B }
\def \plb{  Phys. Lett. B }

\def \prd{  Phys. Rev. D }
\def \prl{  Phys. Rev. Lett.  }

\def \jhep{ J. High Energy Phys. }

\begin{document}

\title{Study of $B_{(s)}^0 \to \phi\phi \to (K^+K^-)(K^+K^-)$ decays in the perturbative QCD approach}

\author{Da-Cheng Yan$^1$}         
\author{Zhou Rui$^2$}              
\author{Zhen-Jun Xiao$^{3}$}    
\author{Ya Li$^4$}                \email[Corresponding author:]{liyakelly@163.com}
\affiliation{$^1$ School of Mathematics and Physics, Changzhou University, Changzhou, Jiangsu 213164, China}
\affiliation{$^2$ College of Sciences, North China University of Science and Technology, Tangshan, Hebei 063210, China}
\affiliation{$^3$ Department of Physics and Institute of Theoretical Physics,
                          Nanjing Normal University, Nanjing, Jiangsu 210023, China}
\affiliation{$^4$ Department of Physics, College of Sciences, Nanjing Agricultural University, Nanjing, Jiangsu 210095, China}
\date{\today}

\begin{abstract}
In this work, we make a detailed analysis on the penguin-dominant processes $B_{(s)}^0 \to \phi\phi \to (K^+K^-)(K^+K^-)$ in the perturbative QCD (PQCD) approach.
In addition to the dominant $P$-wave resonance, the scalar background $f_0(980) \to K^+K^-$ is also accounted for.
We improve the Gegenbauer moments in $KK$ two-meson distribution amplitudes by fitting the PQCD factorization formulas
to measured branching ratios of three-body and four-body $B$ decays.
We extract the branching ratios of two-body $B_{(s)}^0 \to \phi\phi$ decays from the corresponding four-body decay modes and calculate the relevant polarization
fractions together with two relative phases $\phi_{\parallel,\perp}$, which are consistent with the previous theoretical predictions.
The PQCD predictions for the ``true" triple product asymmetries (TPAs) are zero which are expected in the standard model due to the vanishing weak phase difference,
and support the current data reported by the CDF and LHCb Collaborations.
A large ``fake" TPA $\mathcal{A}_\text{T-fake}^1=30.4\%$ of the decay $B^0_s \to \phi\phi \to (K^+K^-)(K^+K^-)$ is predicted for the first time,
which indicates the presence of the significant final-state interactions.
The TPAs of the rare decay channel $B^0 \to \phi\phi\to (K^+K^-)(K^+K^-)$ are also predicted and can be tested in the near future.

\end{abstract}

\maketitle

\section{Introduction}
In the standard model (SM), studies of the polarization amplitudes and triple product asymmetries in the flavour-changing neutral current decays provide powerful tests for the presence of physics beyond the SM~\cite{prd39-3339,npbps13-487,ijmpa19-2505,prd84-096013,plb701-357,prd86-076011,prd88-016007,prd92-076013,prd87-116005}, especially for the decay $B^0_s\to\phi\phi$ via a $b \to s\bar{s}s$ penguin process, where the $\phi(1020)$ is implied throughout the remainder paper.

The $B^0_s\to\phi\phi$ decay is a pseudoscalar to vector-vector transition, where $\phi$ is reconstructed in the $K^+K^-$ final states.
According to the angular momentum conservation, there are three possible spin configurations corresponding to the polarizations of the final-state vector mesons: longitudinal
polarization ($A_0$), and transverse polarization with spins parallel ($A_{\parallel}$)
or perpendicular ($A_{\perp}$) to each other.
The first two states $A_0$ and $A_{\parallel}$ are $CP$ even, while the last one $A_{\perp}$ is $CP$ odd.
Polarization amplitudes can be measured by analyzing angular distributions of final-state particles.
In the factorization assumption, the longitudinal polarization should dominate based on the quark helicity analysis~\cite{zpc1-269,prd64-117503}.
In sharp contrast to these expectations, large transverse polarization of order 50$\%$ is observed in $B\to K^*\phi$, $B\to K^*\rho$ and $B_s\to \phi\phi$ decays~\cite{prl91-201801,prd78-092008,prd85-072005,LHCb:2014xzf,LHCb:2019jgw}, which poses an interesting challenge for the theory.

Interference between the $CP$-even ($A_0$, $A_{\parallel}$) and $CP$-odd ($A_{\perp}$) amplitudes can generate asymmetries in angular distributions, the triple product asymmetries,
which may signal unexpected $CP$ violation due to physics beyond the SM.
In recent years, TPAs have already been measured by Belle, BABAR, CDF and LHCb~\cite{prl95-091601,prd76-031102,prl107-261802,plb713-369,LHCb:2013xyz,LHCb:2014xzf,prd90-052011,jhep07-166,jhep05-026,LHCb:2019jgw}.
These triple products are odd under the time reversal transformation ($T$), and also constitute potential signals of $CP$ violation due to the $CPT$ theorem.
As we know, a non-vanishing direct $CP$ violation needs the interference of at least two amplitudes with a weak phase difference $\Delta \phi$ and a strong phase difference $\Delta \delta$.
The direct $CP$ violation is proportional to $\sin\Delta \phi \sin \Delta \delta$, while TPAs go as $\sin\Delta \phi \cos \Delta \delta$.
The key point is that the direct $CP$ violation can only be produced if there is a nonzero strong phase difference.
It has been argued that all strong phases in $B$ decays should be rather small due to the fact that the $b$-quark is heavy~\cite{ijmpa19-2505}.
Hence, if the strong phases are quite small, the magnitude of the direct $CP$ violation is close to zero, but the TPA is maximal.
It implies that direct $CP$ violation and TPAs complement each other.
Since no tree level operators can contribute to four-body decays $B^0_{(s)} \to \phi\phi \to (K^+K^-)(K^+K^-)$, there is no direct $CP$ violation in such decay modes.
However, $T$-odd triple products (also called ``fake" TPAs), which are proportional to $\cos\Delta \phi \sin \Delta \delta$, can provide useful complementary information.
Thus, it may be more promising to search for TPAs than direct $CP$ asymmetries in $b \to s$ penguin decays.

\begin{figure}[tbp]
	\centering
	\includegraphics[scale=0.7]{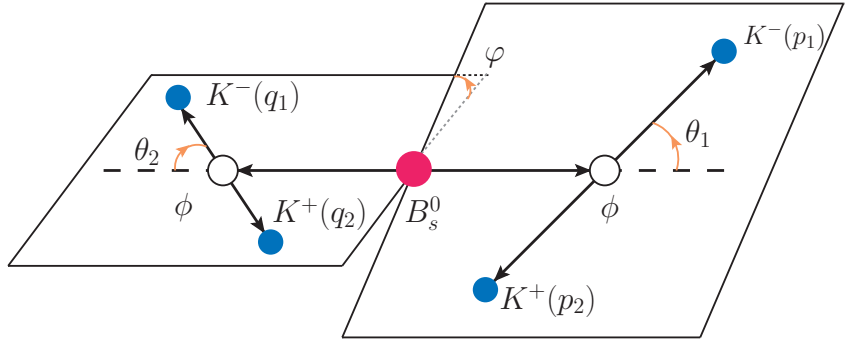}
	\caption{Graphical definitions of the helicity angles $\theta_1$, $\theta_2$ and $\varphi$ for the $B^0_s \to \phi\phi$ decay, with each quasi-two-body intermediate resonance decaying to two pseudoscalars ($\phi \to K^+K^-$).
	$\theta_{1,2}$ is denoted as the angle between the direction of motion of $K^-$ in the $\phi$ rest frame and $\phi$ in the $B^0_s$ rest frame, and $\varphi$ is the angle between the two planes defined by $K^+K^-$ in the $B^0_s$ rest frame.}
\label{fig1}
\end{figure}

$B^0_{(s)} \to \phi\phi$ decays are usually treated as two-body final states on the theoretical sides and have been studied in the two-body framework using QCD factorization (QCDF)~\cite{npb774-64,prd80-114026,2202.08073}, the perturbative QCD approaches~\cite{prd76-074018,prd91-054033,npb935-17}, the soft-collinear effective theory (SCET)~\cite{prd96-073004}  and the factorization-assisted topological amplitude approach (FAT)~\cite{epjc77-333}.
While they are at least four-body decays on the experimental side shown in Fig.~\ref{fig1}, since the vector meson $\phi$ decays via the strong interaction with a nonzero width.
The four-body $B$ meson decays are indeed more challenging than two-body decays, but provide a number of theoretical and phenomenological advantages.
On the one hand, the four-body decay amplitudes depend on five kinematic variables: three helicity angles and two invariant masses of final meson pairs, while the kinematics of two-body decays is fixed.
On the other hand, the four-body decays not only receive the resonant and nonresonant contributions, but also involve the
possible significant final-state interactions (FSIs)~\cite{prd89-094013,1512-09284,prd89-053015}.

Four-body decays are still mostly unexplored from the theoretical point of view since the factorization formalism that describes a multi-body decay in full phase space is not yet available at present.
Recent studies on three-body  hadronic decays of $B$ mesons based on the symmetry principles~\cite{prd72-094031,plb727-136,prd72-075013,prd84-056002,plb728-579,prd91-014029}, the QCDF~\cite{plb622-207,prd74-114009,APPB42-2013,prd76-094006,prd88-114014,prd94-094015,prd89-094007,prd87-076007,jhep10-117,2005-06080,prd99-076010} and the PQCD approaches~\cite{plb763-29,prd95-056008,prd96-093011,prd98-056019,prd98-113003,jpg46-095001,cpc43-073103,epjc79-37,cpc44-073102,jhep03-162,epjc80-394,
2005-02097,prd101-016015,prd102-056017,2105-03899,prd97-033006,epjc77-199} look promising.
It has been proposed that the factorization theorem of three-body $B$ decays is approximately valid when two particles move collinearly and the bachelor particle recoils back~\cite{plb561-258,prd79-094005}.
More details can also be found in Refs.~\cite{1609-07430,npb899-247}.
This situation exists particularly in the low $\pi\pi$ or $K\pi$ invariant mass region ($\lesssim$2 GeV) of the Dalitz plot where most resonant structures are seen.
The Dalitz plot is typically dominated by resonant quasi-two-body contributions along the edge.
This proposal provides a theoretical framework for studies of resonant contributions based on the quasi-two-body-decay mechanism.
Recently, the localized $CP$ violation and branching fraction of the four-body decay
$\bar{B}^0\to K^-\pi^+\pi^+\pi^-$ have been calculated by employing a quasi-two-body QCDF
approach in Refs.~\cite{1912-11874,2008-08458}.
In our previous works~\cite{zjhep,Li:2021qiw,prd105-053002}, the PQCD factorization formalism
based on the quasi-two-body-decay mechanism for four-body $B$ meson decays have been well established.
Within the framework of PQCD approach, the branching ratios and direct $CP$ asymmetries of four-body decays $B_{(s)}^0 \to\pi\pi\pi\pi$ have also been studied~\cite{Liang:2022mrz}.

As a first step, we can only restrict ourselves to the specific kinematical configurations in which each two particles move collinearly and two pairs of final state particles recoil back in the rest frame of the $B$ meson, see Fig.~\ref{fig1}.
Naturally the dynamics associated with the pair of final state mesons can be factorized into a two-meson distribution amplitude (DA)
$\Phi_{h_1h_2}$~\cite{MP,MT01,MT02,MT03,NPB555-231,Grozin01,Grozin02}.
Thereby, the typical PQCD factorization formula for the considered four-body decay amplitude can be described as the form of,
\begin{eqnarray}
\mathcal{A}=\Phi_B\otimes H\otimes \Phi_{KK}\otimes\Phi_{KK},
\end{eqnarray}
where $\Phi_B$ is the universal wave function of the $B$ meson and absorbs the non-perturbative dynamics in the process.
The $\Phi_{KK}$ is the two-hadron DA, which involves the resonant and nonresonant interactions between the two moving collinearly mesons.
The hard kernel $H$ describes the dynamics of the strong and electroweak interactions in four-body hadronic decays in a similar way as the one for the corresponding two-body decays.

\begin{figure}[tbp]
	\centering
	\includegraphics[scale=0.8]{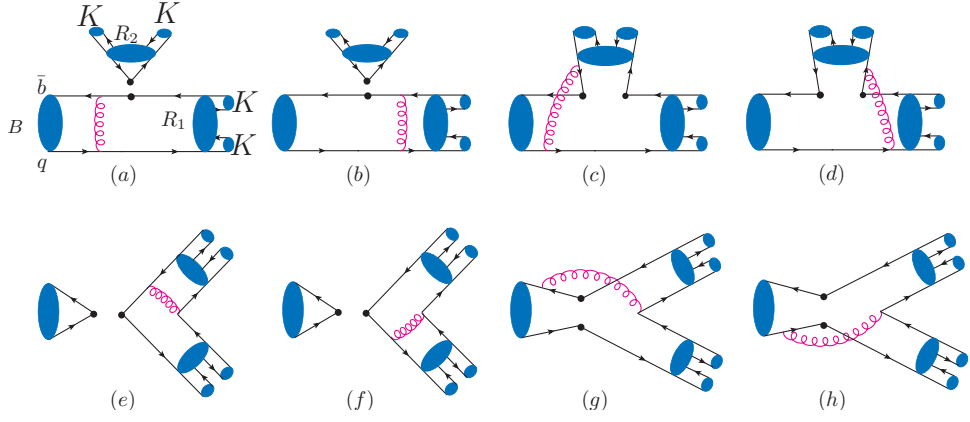}
	\caption{Typical leading-order Feynman diagrams for the four-body decays $B \to (R_1\to) KK (R_2\to) KK$ with $q=(d,s)$, where the symbol $\bullet$ denotes a weak interaction vertex. The diagrams ($a$)-($d$) represent the $B\to (R_1\to) KK$ transition, as well as the diagrams ($e$)-($h$) for annihilation contributions.}
\label{fig2}
\end{figure}

In this work, we study the four-body decays $B^0_{(s)} \to (K^+K^-)(K^+K^-)$ in the PQCD approach based on $k_T$ factorization with the relevant Feynman diagrams illustrated
in Fig.~\ref{fig2}.
The invariant mass of the $K^+K^-$ pair is restricted to be within $\pm30 {\rm MeV}$ of the known mass of the $\phi$ meson for comparison with the LHCb data~\cite{LHCb:2019jgw}.
The effect of identical particles has been considered in our work.
In the considered $(K^+K^-)$ invariant-mass range, the vector resonance $\phi$ is expected to contribute, together with the scalar resonance $f_0(980)$.
The $S$ and $P$-wave contributions are parametrized into the corresponding timelike form factors involved in the two-meson DAs.
We perform a global fit of the Gegenbauer moments in two-kaon DAs associated with both longitudinal and transverse polarizations to measured branching ratios in three-body and four-body charmless hadronic $B$ meson decays, which will be expressed in detail in the following section.
With the improved two-kaon DAs, we calculate the branching ratios and polarization fractions of each partial waves.
In addition, triple-product asymmetries corresponding to the interference of the $CP$-odd amplitudes with the other $CP$-even amplitudes are predicted.

The rest of the paper is organized as follows. The kinematic variables for four-body hadronic
$B$ meson decays are defined in Sec.~\ref{sec:2}.
The considered $S$ and $P$-wave two-meson DAs are also
parametrized, whose normalization form factors are assumed to take the Flatt\'e and relativistic Breit-Wigner (BW) models~\cite{plb63-228,epjc78-1019}.
We explain how to perform the global fit, present and discuss the numerical results in Sec.~\ref{sec:3}, which is followed by the Conclusion.
The Appendix collects the explicit PQCD factorization formulas for all the decay amplitudes.

\section{Framework}\label{sec:2}
\subsection{Kinematics}
Considering the four-body decay $B(p_B)\to R_1(p)R_2(q)\to P_1(p_1)P_2(p_2)Q_1(p_3)Q_2(p_4)$,
as usual, we will work in the $B$ meson rest frame.
By employing the light-cone coordinates, we define the $B$ meson momentum $p_{B}$,
the total momenta of the two kaon-kaon pairs, $p=p_1+p_2$,  $q=p_3+p_4$,
and the quark momentum $k_i$ $(i=B,p,q)$ in each meson in the following form:
\begin{eqnarray}
p_{B}&=&\frac{m_{B}}{\sqrt 2}(1,1,0_{\rm T}),\quad\quad
p=\frac{m_{B}}{\sqrt2}(g^+,g^-,0_{\rm T}),\quad\quad
q=\frac{m_{B}}{\sqrt 2}(f^-,f^+,0_{\rm T}), \nonumber\\
k_{B}&=&\left(0,x_B p_B^+ ,k_{B \rm T}\right),\quad
k_p= \left( x_1 p^+,0,k_{1{\rm T}}\right),\quad\quad
k_q=\left(0,x_2q^-,k_{2{\rm T}}\right),\label{eq:mom-B-k}
\end{eqnarray}
with the $B$ meson mass $m_B$, the parton momentum fractions $x_i$, and the parton transverse momenta $k_{iT}$, $i=B,1,2$.
The explicit expressions of $f^{\pm},g^{\pm}$ related to the invariant masses of the meson pairs via $p^2=\omega_1^2$ and $q^2=\omega_2^2$
can be written as
\begin{eqnarray}\label{eq:epsilon}
g^{\pm}&=&\frac{1}{2}\left[1+\eta_1-\eta_2\pm\sqrt{(1+\eta_1-\eta_2)^2-4\eta_1}\right],\nonumber\\
f^{\pm}&=&\frac{1}{2}\left[1-\eta_1+\eta_2\pm\sqrt{(1+\eta_1-\eta_2)^2-4\eta_1}\right],
\end{eqnarray}
where $\eta_{1,2}=\frac{\omega_{1,2}^2}{m^2_{B}}$.
For the $P$-wave $K K$ pairs, the corresponding longitudinal polarization vectors are defined as
 \begin{eqnarray}\label{eq:pq1}
\epsilon_{p}=\frac{1}{\sqrt{2\eta_1}}(g^+,-g^-,\textbf{0}_{T}),\quad
\epsilon_{q}=\frac{1}{\sqrt{2\eta_2}}(-f^-,f^+,\textbf{0}_{T}),
\end{eqnarray}
which satisfy the normalization $\epsilon_{p}^2=\epsilon_{q}^2=-1$  and the orthogonality
$\epsilon_{p}\cdot p=\epsilon_{q}\cdot q=0$.

The individual momenta $p_i (i=1-4)$ of the four final states can be derived from the relations $p=p_1+p_2$ and $q=p_3+p_4$, together with the on-shell conditions
$p_i^2=m_i^2$ for the final state meson $P_i$ or $Q_i$,
\begin{eqnarray}\label{eq:p1p4}
p_1&=&\left(\frac{m_{B}}{\sqrt{2}}(\zeta_1+\frac{r_1-r_2}{2\eta_1})g^+,\frac{m_{B}}{\sqrt{2}}(1-\zeta_1+\frac{r_1-r_2}{2\eta_1})g^-,\textbf{p}_{T}\right),\nonumber\\
p_2&=&\left(\frac{m_{B}}{\sqrt{2}}(1-\zeta_1-\frac{r_1-r_2}{2\eta_1})g^+,\frac{m_{B}}{\sqrt{2}}(\zeta_1-\frac{r_1-r_2}{2\eta_1})g^-,-\textbf{p}_{T}\right),\nonumber\\
p_3&=&\left(\frac{m_{B}}{\sqrt{2}}(1-\zeta_2+\frac{r_3-r_4}{2\eta_2})f^-,\frac{m_{B}}{\sqrt{2}}(\zeta_2+\frac{r_3-r_4}{2\eta_2})f^+,\textbf{q}_{T}\right),\nonumber\\
p_4&=&\left(\frac{m_{B}}{\sqrt{2}}(\zeta_2-\frac{r_3-r_4}{2\eta_2})f^-,\frac{m_{B}}{\sqrt{2}}(1-\zeta_2-\frac{r_3-r_4}{2\eta_2})f^+,-\textbf{q}_{T}\right),\nonumber\\
\textbf{p}_{\rm T}^2&=&\zeta_1(1-\zeta_1)\omega_1^2+\alpha_1,\quad
\textbf{q}_{\rm T}^2=\zeta_2(1-\zeta_2)\omega_2^2+\alpha_2,
\end{eqnarray}
with the factors
\begin{eqnarray}\label{eq:alpha12}
\alpha_1=-\frac{r_1+r_2}{2\eta_1}+\frac{(r_1-r_2)^2}{4\eta_1^2},\quad
\alpha_2=-\frac{r_3+r_4}{2\eta_2}+\frac{(r_3-r_4)^2}{4\eta_2^2},
\end{eqnarray}
and the mass ratios $r_i=m_i^2/m^2_B$, $m_i$ being the masses of the final state mesons.

Comparing Eqs.~(\ref{eq:p1p4}) and~(\ref{eq:mom-B-k}), one can see that the meson momentum fractions are modified by the final state meson masses,
\begin{eqnarray}\label{eq:momfrac}
\frac{p_1^+}{p^+}=\zeta_1+\frac{r_1-r_2}{2\eta_1},\quad \frac{q_1^-}{q^-}=\zeta_2+\frac{r_3-r_4}{2\eta_2}.
\end{eqnarray}

The relation between $\zeta_{1,2}$ and the polar angle $\theta_{1,2}$ in the dimeson rest frame in Fig.~\ref{fig1} can be obtained easily,
\begin{eqnarray}\label{eq:cos}
2\zeta_{1}-1=\sqrt{1+4\alpha_1}\cos\theta_1, ~\quad
2\zeta_{2}-1=\sqrt{1+4\alpha_2}\cos\theta_2,
\end{eqnarray}
with the upper and lower limits of $\zeta_{1,2}$
\begin{eqnarray}
\zeta_{1\text{max,min}}=\frac{1}{2}\left[1\pm\sqrt{1+4\alpha_1}\right],~\quad
\zeta_{2\text{max,min}}=\frac{1}{2}\left[1\pm\sqrt{1+4\alpha_2}\right].
\end{eqnarray}
\subsection{Distribution amplitudes}
Without the endpoint singularities in the evaluations,
the distribution amplitudes are one of the most significant nonperturbative inputs in the PQCD approach.
In this section, we will briefly introduce the $B$ meson DAs, the $S$-, $P$-wave two-kaon DAs, as well as the time-like form factors used in our calculations.
In what follows the subscripts $S,P$ are always related to the corresponding partial waves.

The light-cone hadronic matrix element for a $B$ meson is parametrized
as~\cite{prd63-054008,prd65-014007,epjc28-515,ppnp51-85,Prd85-094003,fop2021-16}
\begin{eqnarray}
\int d^4z e^{i{\bf k_1}\cdot z}\langle0|q_{\beta}(z)\bar{b}_{\alpha}(0)|B(p_{B})\rangle=
\frac{i}{\sqrt{2N_c}}\left\{({p \hspace{-2.0truemm}/}_{B}+m_{B})\gamma_{5}\left[\phi_{B}({\bf k_1})-
\frac{{n \hspace{-2.0truemm}/}-{v \hspace{-2.0truemm}/}}{\sqrt{2}}\bar{\phi}_{B}({\bf k_1})\right]\right\}
_{\beta\alpha},
\end{eqnarray}
where $q$ represents a $d$ or $s$ quark. The two wave functions $\phi_{B}$ and
$\bar \phi_{B}$ in the above decomposition, related to $\phi_{B}^+$ and $\phi_{B}^-$ defined in the
literature~\cite{GN} via $\phi_{B}=(\phi_{B}^++\phi_{B}^-)/2$ and $\bar \phi_{B}=(\phi_{B}^+-\phi_{B}^-)/2$,
obey the normalization conditions
\begin{eqnarray}
\int \frac{d^4{\bf k_1}}{(2\pi)^4}\phi_{B}({\bf k_1})=\frac{f_{B}}{2\sqrt{2N_c}},\;\;\;\;
\int\frac{d^4{\bf k_1}}{(2\pi)^4}\bar{\phi}_{B}({\bf k_1})=0.
\end{eqnarray}
It has been shown that the contribution from $\bar \phi_{B}$ is of next-to-leading power
and numerically suppressed~\cite{prd65-014007,epjc28-515,Prd103-056006}, compared to the leading-power contribution from $\phi_B$.
Taking the PQCD evaluation of the  $B\to\pi$ transition form factor $F_0^{B\to \pi}$ in Ref.~\cite{Prd103-056006}
as an example, we find that
the  $\bar{\phi}_{B}$ contribution to $F_0^{B\to \pi}$ is about 20\% of the $\phi_B$ one.
The higher-twist $B$ meson DAs have been
systematically investigated in the heavy quark effective theory~\cite{jhep05-022},
which are decomposed according to definite twist and conformal spin assignments up to twist 6.
In principle, all the next-to-leading-power sources should be included
for a consistent and complete analysis, which, however, goes beyond the scope of the present formalism.
Therefore, we focus only on the leading-power component
\begin{eqnarray}
\Phi_B= \frac{i}{\sqrt{2N_c}} ({ p \hspace{-2.0truemm}/ }_B +m_B) \gamma_5 \phi_B ( x,b), \label{bmeson}
\end{eqnarray}
with the impact parameter $b$ being  conjugate to the parton transverse momentum $k_{B \rm T}$.
The $B$ meson DA $\phi_B (x,b)$ is chosen as the model form widely adopted in
the PQCD approach~\cite{prd63-054008,prd65-014007,epjc28-515,ppnp51-85,Prd85-094003,Li:2012md},
\begin{eqnarray}
\phi_B(x,b)&=& N_B x^2(1-x)^2\mathrm{exp} \left  [ -\frac{m_B^2 x^2}{2 \omega_{B}^2} -\frac{1}{2} (\omega_{B} b)^2\right] ,
 \label{phib}
\end{eqnarray}
where the constant $N_B$ is related to the $B$ meson decay constant $f_B$
through the normalization condition $\int_0^1dx \; \phi_B(x,b=0)=f_B/(2\sqrt{2N_c})$.
The shape parameter takes the values $\omega_B = 0.40$ GeV for $B^0$ meson and $\omega_{B_s}=0.48$
GeV~\cite{prd63-054008,plb504-6,prd63-074009,2012-15074} for $B^0_s$ meson with 10\% variation in the
numerical study below.

The $S$-wave two-kaon DA can be written in the following form~\cite{prd91-094024},
\begin{eqnarray}\label{swave}
\Phi_{S}(z,\omega)=\frac{1}{\sqrt{2N_c}}[{p\hspace{-1.5truemm}/}\phi^0_S(z,\omega^2)+
\omega\phi^s_S(z,\omega^2)+\omega({n\hspace{-2.0truemm}/}{v\hspace{-2.0truemm}/}-1)\phi^t_S(z,\omega^2)],
\end{eqnarray}
in which the asymptotic forms of the individual twist-2 and twist-3 components $\phi_S^0$, $\phi_S^{s,t}$
are parametrized as~\cite{MP,MT01,MT02,MT03}
\begin{eqnarray}
\phi^0_S(z,\omega^2)&=&\frac{9F_S(\omega^2)}{\sqrt{2N_c}}a_S z(1-z)(1-2z),\label{eq:phis0}\\
\phi^s_S(z,\omega^2)&=&\frac{F_S(\omega^2)}{2\sqrt{2N_c}},\\
\phi^t_S(z,\omega^2)&=&\frac{F_S(\omega^2)}{2\sqrt{2N_c}}(1-2z),
\end{eqnarray}
with the time-like scalar form factor $F_S(\omega^2)$.
The Gegenbauer moment $a_S$ in Eq.~(\ref{eq:phis0}) is adopted the same value as that determined in Ref.~\cite{epjc79-792}: $a_S=0.80\pm 0.16$.

The corresponding $P$-wave two-kaon DAs related to both longitudinal and transverse polarizations are decomposed,
up to the twist 3, into~\cite{prd98-113003}:
\begin{eqnarray}
\Phi_P^{L}(z,\zeta,\omega)&=&\frac{1}{\sqrt{2N_c}} \left [{ \omega \epsilon\hspace{-1.5truemm}/_p  }\phi_P^0(z,\omega^2)+\omega\phi_P^s(z,\omega^2)
+\frac{{p\hspace{-1.5truemm}/}_1{p\hspace{-1.5truemm}/}_2
  -{p\hspace{-1.5truemm}/}_2{p\hspace{-1.5truemm}/}_1}{\omega(2\zeta-1)}\phi_P^t(z,\omega^2) \right ] (2\zeta-1)\;,\label{pwavel}\\
\Phi_P^{T}(z,\zeta,\omega)&=&\frac{1}{\sqrt{2N_c}}
\Big [\gamma_5{\epsilon\hspace{-1.5truemm}/}_{T}{ p \hspace{-1.5truemm}/ } \phi_P^T(z,\omega^2)
+\omega \gamma_5{\epsilon\hspace{-1.5truemm}/}_{T} \phi_P^a(z,\omega^2)+ i\omega\frac{\epsilon^{\mu\nu\rho\sigma}\gamma_{\mu}
\epsilon_{T\nu}p_{\rho}n_{-\sigma}}{p\cdot n_-} \phi_P^v(z,\omega^2) \Big ]\non
&&\cdot \sqrt{\zeta(1-\zeta)+\alpha_1}\label{pwavet}\;.
\end{eqnarray}
The various twists $\phi_P^i$ in the above equations can be expanded in terms of the Gegenbauer polynomials:
\begin{eqnarray}
\phi_P^0(z,\omega^2)&=&\frac{3F_P^{\parallel}(\omega^2)}{\sqrt{2N_c}}z(1-z)\left[1
+a^0_{2\phi}\frac{3}{2}(5(1-2z)^2-1)\right]\label{eq:phi0} \;,\\
\phi_P^s(z,\omega^2)&=&\frac{3F_P^{\perp}(\omega^2)}{2\sqrt{2N_c}}(1-2z)\;,\\
\phi_P^t(z,\omega^2)&=&\frac{3F_P^{\perp}(\omega^2)}{2\sqrt{2N_c}}(1-2z)^2 \;,\\
\phi_P^T(z,\omega^2)&=&\frac{3F_P^{\perp}(\omega^2)}
{\sqrt{2N_c}}z(1-z)[1+a^{T}_{2\phi}\frac{3}{2}(5(1-2z)^2-1)]\;,\\
\label{eq:phiT}
\phi_P^a(z,\omega^2)&=&\frac{3F_P^{\parallel}(\omega^2)}
{4\sqrt{2N_c}}(1-2z)\;,\\
\phi_P^v(z,\omega^2)&=&\frac{3F_P^{\parallel}(\omega^2)}
{8\sqrt{2N_c}}[1+(1-2z)^2]\;\label{eq:phiv},
\end{eqnarray}
with the Gegenbauer coefficients $a_{2\phi}^{0,T}$ and the two $P$-wave form factors $F_P^{\parallel}(\omega^2)$ and $F_P^{\perp}(\omega^2)$.
The moment $a_{2\phi}^{0}$ in the longitudinal twist-2 component $\phi_P^{0}$ has already been
determined in a recent global analysis from the three-body $B$ decays in the PQCD approach~\cite{2105-03899} .
We will update the fitting result in the following section by taking
the additional four-body decay $B_s^0\to \phi\phi\to (K^+K^-)(K^+K^-)$ into account,
while the moment $a_{2\phi}^{T}$ associated with the transverse twist-2 component $\phi_P^{T}$ is determined in a global analysis for the first time.
Since the amounts of the current experimental data are not yet enough for fixing the Gegenbauer moments in the twist-3 DAs $\phi_P^{s,t}$ and $\phi_P^{v,a}$,
they have been set to the asymptotic forms in the present work.

The elastic rescattering effects in a final-state meson pair can be absorbed into the time-like form factors
$F^{\parallel,\perp}_P(\omega^2)$ in the two-meson DAs according to the Watson theorem~\cite{pr88-1163}.
For the narrow resonance $\phi$, we usually employ the relativistic BW line shape for the form factors $F_P^{\parallel}(\omega^2)$~\cite{BW-model}.
The explicit formula is expressed as~\cite{epjc78-1019}
\begin{eqnarray}
\label{BRW}
F_P^{\parallel}(\omega^2)&=&\frac{ m_{\phi}^2}{m^2_{\phi} -\omega^2-im_{\phi}\Gamma_{\phi}(\omega^2)} \;,
\end{eqnarray}
where $m_{\phi}=1.0195$ GeV is the $\phi$ meson mass.
The mass-dependent width $\Gamma_{\phi}(\omega)$ is defined as
\begin{eqnarray}
\label{BRWl}
\Gamma_{\phi}(\omega^2)&=&\Gamma_{\phi}\left(\frac{m_{\phi}}{\omega}\right)\left(\frac{k(\omega)}{k(m_{\phi})}\right)^{(2L_R+1)},
\end{eqnarray}
with the natural width of the $\phi$ meson $\Gamma_{\phi}=4.25$ MeV~\cite{pdg2020}.
The $k(\omega)$ is the momentum vector of the resonance decay product measured in the resonance rest frame, while $k(m_{\phi})$
is the value of $k(\omega)$ when $\omega=m_{\phi}$.
The explicit expression of kinematic variables $k(\omega)$ is defined in the $h_1h_2$ center-of-mass frame
\begin{eqnarray}
k(\omega)=\frac{\sqrt{\lambda(\omega^2,m_{h_1}^2,m_{h_2}^2)}}{2\omega},
\end{eqnarray}
with the K$\ddot{a}$ll$\acute{e}$n function $\lambda (a,b,c)= a^2+b^2+c^2-2(ab+ac+bc)$ and $m_{h_1,h_2}$ being the final state mass.
The orbital angular momentum $L_R$ in the two-meson system is set to $L_R = 1$ for a $P$-wave state.
Due to the limited studies on the form factor $F_P^{\perp}(\omega^2)$, we use the two decay
constants $f_{\phi}^{(T)}$ of the intermediate particle to determine the ratio $F_P^{\perp}(\omega^2)/F_P^{\parallel}(\omega^2)\approx (f_{\phi}^T/f_{\phi})$.

For scalar resonance $f_0(980)$, we adopt the Flatt\'e parametrization where the resulting line shape is above and below the threshold of the intermediate particle~\cite{plb63-228}.
If the coupling of a resonance to the channel opening nearby is very strong, the Flatt\'e parametrization shows a scaling invariance and does not allow for an extraction of individual partial decay widths.
Thus, we employ the modified Flatt\'e model suggested by D.V.~Bugg~\cite{prd78-074023} following the LHCb collaboration~\cite{prd89-092006,prd90-012003},
\begin{eqnarray}
F_S(\omega^2)=\frac{m_{f_0(980)}^2}{m_{f_0(980)}^2-\omega^2-im_{f_0(980)}(g_{\pi\pi}\rho_{\pi\pi}+g_{KK}\rho_{KK}F^2_{KK})}\;.
\end{eqnarray}
The coupling constants $g_{\pi\pi}=0.167$ GeV and $g_{KK}=3.47g_{\pi\pi}$~\cite{prd89-092006,prd90-012003}
describe the $f_0(980)$ decay into the final states $\pi^+\pi^-$ and $K^+K^-$, respectively.
The exponential factor $F_{KK}=e^{-\alpha q_K^2}$
is introduced above the $K\bar{K}$ threshold to reduce the $\rho_{KK}$ factor as invariant mass increases,
where $q_k$ is the momentum of the kaon in the $K\bar{K}$ rest frame and $\alpha=2.0\pm 0.25$ GeV$^{-2}$~\cite{prd89-092006,prd78-074023}.
The phase space factors $\rho_{\pi\pi}$ and $\rho_{KK}$ read
as~\cite{prd87-052001,prd89-092006,plb63-228}
\begin{eqnarray}
\rho_{\pi\pi}=\frac23\sqrt{1-\frac{4m^2_{\pi^\pm}}{\omega^2}}
 +\frac13\sqrt{1-\frac{4m^2_{\pi^0}}{\omega^2}},\quad
\rho_{KK}=\frac12\sqrt{1-\frac{4m^2_{K^\pm}}{\omega^2}}
 +\frac12\sqrt{1-\frac{4m^2_{K^0}}{\omega^2}}.
\end{eqnarray}

\subsection{Helicity amplitudes}
The differential branching fraction for the $B^0_{(s)}\rightarrow (K^+K^-)(K^+K^-)$
in the $B^0_{(s)}$ meson rest frame is expressed as
\begin{eqnarray}\label{eq:decayrate}
\frac{d^5\mathcal{B}}{d\Omega}=\frac{\tau_B k(\omega_1)k(\omega_2)k(\omega_1,\omega_2)}{16(2\pi)^6m_B^2} |A|^2, 
\end{eqnarray}
where $d \Omega$ with $\Omega\equiv \{\theta_1, \theta_2,\varphi,\omega_1,\omega_2\}$ stands for the
five-dimensional measure spanned by the three helicity angles and the two invariant masses, and
\begin{eqnarray}
k(\omega_1,\omega_2)&=&\frac{\sqrt{[m_B^2-(\omega_1+\omega_2)^2][m_B^2-(\omega_1-\omega_2)^2]}}{2m_B},
\end{eqnarray}
is the momentum of the $K^+K^-$ pair in the $B_{(s)}$ meson rest frame.

The four-body phase space has been derived in the analysis of the
$K\rightarrow \pi\pi l\nu$ decay~\cite{pr168-1858}, the
semileptonic $\bar{B}\rightarrow D(D^*)\pi l \nu$ decays~\cite{prd48-3204},
semileptonic baryonic decays~\cite{prd85-094019,plb780-100}, and four-body baryonic decays~\cite{plb770-348}.
One can confirm that Eq.~(\ref{eq:decayrate}) is equivalent to those in Refs.~\cite{prd85-094019,plb770-348}
by appropriate variable changes.
Replacing the helicity angle $\theta$ by the meson momentum fraction $\zeta$ via Eq.~(\ref{eq:cos}),
the Eq.~(\ref{eq:decayrate}) is turned into
\begin{eqnarray}\label{eq:decayrate1}
\frac{d^5\mathcal{B}}{d\zeta_1d\zeta_2d \omega_1d \omega_2d\varphi}=
\frac{\tau_B k(\omega_1)k(\omega_2)k(\omega_1,\omega_2)}{4(2\pi)^6m_B^2\sqrt{1+4\alpha_1}\sqrt{1+4\alpha_2}}|A|^2.
\end{eqnarray}

The $B_s^0 \to \phi\phi \to (K^+K^-)(K^+K^-)$ decay comprises a mixture of $CP$ eigenstates and can be disentangled by means of an angular analysis in the helicity basis.
In this basis, the decay is described by three angles $\theta_1$, $\theta_2$ and $\varphi$, depicted in Fig.~\ref{fig1}, where the $\theta_{1,2}$ is the angle between the $K^-$ direction in the $\phi$ rest frame and the $\phi$ direction in the $B^0_s$ rest frame, and $\varphi$ is the angle between the two $\phi$ meson decay planes.

Due to the proximity of the $\phi$ resonance to the scalar $f_0(980)$ resonance, there are irreducible scalar resonant
contributions to four-body $B_{(s)}^0 \to (K^+K^-)(K^+K^-)$ decays.
Thereby, a $K^+K^-$ pair can be produced in the $S$ or $P$-wave configuration in the selected invariant mass regions.
One decomposes the decay amplitudes into six helicity components: $h=VV$ (3), $VS$ (2), and $SS$, each with a corresponding amplitude $A_h$, where $V$ denotes a vector meson and $S$ denotes a scalar meson.
The first three, commonly referred to as the $P$-wave amplitudes, are associated with
the final states, where both $K^+K^-$ pairs come from intermediate vector mesons.
In the transversity basis, a $P$-wave decay amplitude can be decomposed into three components:
$A_0$, for which the polarizations of the final-state vector mesons are
longitudinal to their momenta, and $A_{\parallel}$ ($A_\perp$), for which the polarizations
are transverse to the momenta and parallel (perpendicular) to each other.
As the $S$-wave $K^+K^-$ pair can arise from $R_1$ or $R_2$ labelled in Fig.~\ref{fig2}(a),
the corresponding single $S$-wave amplitude is denoted $A_{VS}$.
The double $S$-wave amplitude $A_{SS}$ is associated with the final state,
where both two-meson pairs are generated in the $S$ wave.
A randomised choice is made for which $\phi$ meson is used to determine $\theta_1$ and which is used to determine $\theta_2$.
Thus, the total decay amplitude $A$ is a coherent sum of the $P$-, $S$-, and double $S$-wave components.
Specifically, these helicity amplitudes for the $B^0_{(s)}\to (K^+K^-)(K^+K^-)$ decays
denote
\begin{eqnarray}
A_{VV}&:& B_{(s)}^0 \rightarrow \phi(\to K^+K^-) \phi(\to K^+K^-), \nonumber\\
A_{VS}&:& B_{(s)}^0 \rightarrow \phi(\to K^+K^-)f_0(980)(\to K^+K^-) ,\nonumber\\
A_{SS}&:& B_{(s)}^0 \rightarrow f_0(980)(\to K^+K^-) f_0(980)(\to K^+K^-).
\end{eqnarray}

By including the $\zeta_{1,2}$ dependencies instead of $\theta_{1,2}$ and azimuth-angle dependencies relying on  Eq.~(\ref{eq:cos}), the total decay amplitude in Eq.~(\ref{eq:decayrate1}) can be written as
\begin{eqnarray}\label{eq:allampli}
A&=&A_0\frac{2\zeta_1-1}{\sqrt{1+4\alpha_1}}\frac{2\zeta_2-1}{\sqrt{1+4\alpha_2}}
+A_{\parallel}2\sqrt{2}\sqrt{\frac{\zeta_1(1-\zeta_1)+\alpha_1}{1+4\alpha_1}}
\sqrt{\frac{\zeta_2(1-\zeta_2)+\alpha_2}{1+4\alpha_2}}\cos\varphi \nonumber\\
&&
+i A_{\perp}2\sqrt{2}\sqrt{\frac{\zeta_1(1-\zeta_1)+\alpha_1}{1+4\alpha_1}}
\sqrt{\frac{\zeta_2(1-\zeta_2)+\alpha_2}{1+4\alpha_2}}\sin\varphi\non
&& +A_{VS}(\frac{2\zeta_1-1}{\sqrt{1+4\alpha_1}}+\frac{2\zeta_2-1}{\sqrt{1+4\alpha_2}})+A_{SS}.
\end{eqnarray}

On basis of Eq.~(\ref{eq:decayrate1}), we can obtain the branching ratio form,
\begin{eqnarray}\label{eq:brsss}
\mathcal{B}_h=\frac{\tau_B}{4(2\pi)^6m_B^2}\frac{2\pi}{9}C_h
\int d\omega_1d\omega_2k(\omega_1)k(\omega_2)k(\omega_1,\omega_2)|A_h|^2,
\end{eqnarray}
where the invariant masses $\omega_{1,2}$ are integrated over the chosen $K^+K^-$ mass window.
The coefficients $C_h$ are the results of the integrations over $\zeta_1,\zeta_2,\varphi$ in terms of Eq.~(\ref{eq:brsss}) and listed as follows,
\begin{eqnarray}\label{eq:radii}
C_h=\left\{\begin{aligned}
&(1+4\alpha_1)(1+4\alpha_2), \quad\quad\quad  &h=0,\parallel,\perp, \\
&3(1+4\alpha_{1}) ,\quad\quad\quad  &h=VS, \\
&9, \quad\quad\quad  &h=SS. \\
\end{aligned}\right.
\end{eqnarray}

The $CP$-averaged branching ratio and the direct $CP$ asymmetry in each component are defined as below,
\begin{eqnarray}
\mathcal{B}_h^{\rm avg}=\frac{1}{2}(\mathcal{B}_h+\mathcal{\bar{B}}_h),
\quad \mathcal{A}^{\text{dir}}_h=\frac{\mathcal{\bar{B}}_h-\mathcal{B}_h}{\mathcal{\bar{B}}_h+\mathcal{B}_h},
\end{eqnarray}
respectively, where $\mathcal{\bar{B}}_h$ is the branching ratio of the corresponding $CP$-conjugate channel.
The sum of the six components yields the total branching ratio and the overall direct-$CP$ asymmetry,
\begin{eqnarray}
\mathcal{B}_{\text{total}}=\sum_h \mathcal{B}_h,\quad
\mathcal{A}^{\text{dir}}_{CP}=\frac{\sum_h \mathcal{\bar{B}}_h
-\sum_h \mathcal{B}_h}{\sum_h \mathcal{\bar{B}}_h+\sum_h \mathcal{B}_h},
\end{eqnarray}
respectively.

For the $VV$ decays, the polarization fractions $f_{\lambda}$ with $\lambda=0$, $\parallel$,
and $\perp$ and two relative phases $\phi_{\parallel}$, $\phi_{\perp}$ are described as
\begin{eqnarray}\label{pol}
f_{\lambda}=\frac{|A_{\lambda}|^2}{|A_0|^2
+|A_{\parallel}|^2+|A_{\perp}|^2}, \quad \phi_{\parallel,\perp}=\text{arg}\frac{A_{\parallel,\perp}}{A_0},
\end{eqnarray}
with the normalisation relation $f_0+f_{\parallel}+f_{\perp}=1$.

\subsection{Triple product asymmetries}\label{sec:TPAs}
Consider a four-body decay $B \to R_1 (\to P_1P_2) R_2(\to Q_1Q_2)$, in which one measures the four particles' momenta in the $B$ rest frame.
We define $\hat{n}_{R_i} (i=1,2)$ is a unit vector perpendicular to the $R_i$ decay plane and $\hat{z}_{R_1}$ is a unit vector in the direction of $R_1$ in the $B$ rest frame.
Thus we have
\begin{eqnarray}\label{TP1}
\hat{n}_{R_1}\cdot \hat{n}_{R_2}=\cos\varphi, ~\quad \hat{n}_{R_1}\times \hat{n}_{R_2}=\sin\varphi \hat{z},
\end{eqnarray}
implying a $T$-odd scalar triple product
\begin{eqnarray}\label{TP2}
(\hat{n}_{R_1}\times \hat{n}_{R_2})\cdot \hat{z}=\sin\varphi, ~\quad 2(\hat{n}_{R_1}\cdot \hat{n}_{R_2})(\hat{n}_{R_1}\times \hat{n}_{R_2})\cdot\hat{z}=\sin2\varphi .
\end{eqnarray}

One can define a TPA as an asymmetry between the number of decays involving positive and negative values of $\sin\varphi$ or $\sin2\varphi$,
\begin{eqnarray} \label{eq:ATs}
\mathcal{A}_{\text{T}}^1&=&\frac{\Gamma(\cos\theta_1\cos\theta_2\sin\varphi>0)-\Gamma(\cos\theta_1\cos\theta_2\sin\varphi<0)}
{\Gamma(\cos\theta_1\cos\theta_2\sin\varphi>0)+\Gamma(\cos\theta_1\cos\theta_2\sin\varphi<0)},\\
\mathcal{A}_{\text{T}}^2&=&\frac{\Gamma(\sin2\varphi>0)-\Gamma(\sin2\varphi<0)}
{\Gamma(\sin2\varphi>0)+\Gamma(\sin2\varphi<0)}.
\end{eqnarray}

It has been found that TPAs originate from the interference of the $CP$-odd amplitudes $A_{\perp}$ with the other $CP$-even amplitudes $A_0$, $A_{\parallel}$.
According to Eq.~(\ref{eq:cos}), the TPAs associated with $A_{\perp}$ for the considered four-body decays are derived from the
partially integrated differential decay rates as~\cite{prd84-096013,jhep07-166}
\begin{eqnarray} \label{eq:ATs2}
\mathcal{A}_{\text{T}}^1&=&\frac{\Gamma((2\zeta_1-1)(2\zeta_2-1)\sin\varphi>0)-\Gamma((2\zeta_1-1)(2\zeta_2-1)\sin\varphi<0)}
{\Gamma((2\zeta_1-1)(2\zeta_2-1)\sin\varphi>0)+\Gamma((2\zeta_1-1)(2\zeta_2-1)\sin\varphi<0)}\nonumber\\&=&
-\frac{2\sqrt{2}}{\pi\mathcal{D}}\int d\omega_1 d\omega_2k(\omega_1)k(\omega_2)k(\omega_1,\omega_2) \text{Im}[A_{\perp}A_0^*],\\
\mathcal{A}_{\text{T}}^2&=&\frac{\Gamma(\sin2\varphi>0)-\Gamma(\sin2\varphi<0)}
{\Gamma(\sin2\varphi>0)+\Gamma(\sin2\varphi<0)}\nonumber\\&=&
-\frac{4}{\pi\mathcal{D}}\int d\omega_1 d\omega_2k(\omega_1)k(\omega_2)k(\omega_1,\omega_2) \text{Im}[A_{\perp}A_{\parallel}^*],
\end{eqnarray}
with the denominator
\begin{eqnarray}
\mathcal{D}=\int d\omega_1 d\omega_2 k(\omega_1)k(\omega_2)k(\omega_1,\omega_2)(|A_0|^2+|A_{\parallel}|^2+|A_{\perp}|^2).
\end{eqnarray}
The above TPAs contain the integrands $\text{Im}(A_{\perp}A_{0,\parallel}^*)=|A_{\perp}||A_{0,\parallel}^*|\sin(\Delta\phi+\Delta \delta)$,
where $\Delta\phi$ and $\Delta\delta$ denote the weak and strong phase differences between the amplitudes
$A_{\perp}$ and $A_{0,\parallel}$, respectively.
As already noted, $\text{Im}(A_{\perp}A_{0,\parallel}^*)$ can be nonzero even if the weak phases vanish.
Thus, it is not quite accurate to identify a nonzero TPA as a signal of $CP$ violation.
To obtain a true $CP$ violation signal, one has to compare the TPAs in the $B$ and $\bar{B}$ meson decays.
The helicity amplitude for the $CP$-conjugated process can be inferred from  Eq.~(\ref{eq:allampli})
through $A_0\to \bar{A}_0$, $A_{\parallel} \to \bar{A}_{\parallel}$ and $A_{\perp} \to -\bar{A}_{\perp}$, in which the $\bar{A}_{\lambda}$ are obtained from the $A_{\lambda}$ by changing the sign of the weak phases.
Thus, the TPAs $\bar{\mathcal{A}}_{\text{T}}^i$ for the charge-conjugate process are defined similarly,
but with a multiplicative minus sign.

One therefore constructs the ``true" and ``fake" asymmetries
by combining $\mathcal{A}_\text{T}^i$ and $\bar{\mathcal{A}}_\text{T}^i$~\cite{prd84-096013}
\begin{eqnarray}\label{eq:tpa}
\mathcal{A}_\text{T-true}^1&\equiv& \frac{[\Gamma(T>0)+\bar{\Gamma}(\bar{T}>0)]-[\Gamma(T<0)+\bar{\Gamma}(\bar{T}<0)]}
{[\Gamma(T>0)+\bar{\Gamma}(\bar{T}>0)]+[\Gamma(T<0)+\bar{\Gamma}(\bar{T}<0)]}\non
&=&-\frac{2\sqrt{2}}{\pi(\mathcal{D}+\bar{\mathcal{D}})}\int d\omega_1 d\omega_2k(\omega_1)k(\omega_2)k(\omega_1,\omega_2) \text{Im}[A_{\perp}A_0^*-\bar{A}_{\perp}\bar{A}_0^*], \label{tp1-t}\\
\mathcal{A}_\text{T-true}^2&\equiv& \frac{[\Gamma(\sin2\varphi>0)+\bar{\Gamma}(\sin2\bar{\varphi}>0)]-[\Gamma(\sin2\varphi<0)+\bar{\Gamma}(\sin2\bar{\varphi}<0)]}
{[\Gamma(\sin2\varphi>0)+\bar{\Gamma}(\sin2\bar{\varphi}>0)]+[\Gamma(\sin2\varphi<0)+\bar{\Gamma}(\sin2\bar{\varphi}<0)]}\non
&=&-\frac{4}{\pi(\mathcal{D}+\bar{\mathcal{D}})}\int d\omega_1 d\omega_2k(\omega_1)k(\omega_2)k(\omega_1,\omega_2) \text{Im}[A_{\perp}A_{\parallel}^*-\bar{A}_{\perp}\bar{A}_{\parallel}^*], \label{tp2-t}\\
\mathcal{A}_\text{T-fake}^1&\equiv& \frac{[\Gamma(T>0)-\bar{\Gamma}(\bar{T}>0)]-[\Gamma(T<0)-\bar{\Gamma}(\bar{T}<0)]}
{[\Gamma(T>0)+\bar{\Gamma}(\bar{T}>0)]+[\Gamma(T<0)+\bar{\Gamma}(\bar{T}<0)]}\non
&=&-\frac{2\sqrt{2}}{\pi(\mathcal{D}+\bar{\mathcal{D}})}\int d\omega_1 d\omega_2k(\omega_1)k(\omega_2)k(\omega_1,\omega_2) \text{Im}[A_{\perp}A_0^*+\bar{A}_{\perp}\bar{A}_0^*], \label{tp1-f}\\
\mathcal{A}_\text{T-fake}^2&\equiv& \frac{[\Gamma(\sin2\varphi>0)-\bar{\Gamma}(\sin2\bar{\varphi}>0)]-[\Gamma(\sin2\varphi<0)-\bar{\Gamma}(\sin2\bar{\varphi}<0)]}
{[\Gamma(\sin2\varphi>0)+\bar{\Gamma}(\sin2\bar{\varphi}>0)]+[\Gamma(\sin2\varphi<0)+\bar{\Gamma}(\sin2\bar{\varphi}<0)]}\non
&=&-\frac{4}{\pi(\mathcal{D}+\bar{\mathcal{D}})}\int d\omega_1 d\omega_2k(\omega_1)k(\omega_2)k(\omega_1,\omega_2) \text{Im}[A_{\perp}A_{\parallel}^*+\bar{A}_{\perp}\bar{A}_{\parallel}^*], \label{tp2-f}
\end{eqnarray}
with $\bar{\Gamma}$ being the decay rate of the $CP$-conjugate process, $T=(2\zeta_1-1)(2\zeta_2-1)\sin\varphi$ and $\bar{T}=(2\zeta_1-1)(2\zeta_2-1)\sin\bar{\varphi}$ and the denominator is
\begin{eqnarray}
\bar{\mathcal{D}}=\int d\omega_1 d\omega_2 k(\omega_1)k(\omega_2)k(\omega_1,\omega_2)(|\bar{A}_0|^2+|\bar{A}_{\parallel}|^2+|\bar{A}_{\perp}|^2),
\end{eqnarray}
for the $CP$-conjugate decay.

It is shown that the terms $\text{Im}[A_{\perp}A_{0,\parallel}^*-\bar{A}_{\perp}\bar{A}_{0,\parallel}^*]$ in Eqs.~(\ref{tp1-t}) and (\ref{tp2-t}) are proportional to $\sin\Delta\phi\cos\Delta\delta$, which are nonzero only in the presence of the weak phase difference.
Then TPAs provide an alternative measure of $CP$ violation.
Furthermore, compared with direct $CP$ asymmetries, $\mathcal{A}_\text{T-true}^i$ does not suffer the suppression from the strong phase difference, and is maximal when the strong phase difference vanishes.
For the special case of the involved neutral intermediate states $B^0_{(s)}\rightarrow \phi\phi$ modes, in which each helicity amplitude involves the same single weak phase in the SM.
This results in $\mathcal{A}_\text{T-true}^i=0$  due to the vanishing weak phase difference.
The ``true" TPAs for the neutral modes are thus expected to be zero in the SM.
If such asymmetries are observed experimentally, it is probably a signal of new physics.
While for the term $\text{Im}[A_{\perp}A_{0,\parallel}^*+\bar{A}_{\perp}\bar{A}_{0,\parallel}^*]\propto\cos\Delta\phi\sin\Delta\delta$,
the $\mathcal{A}_\text{T-fake}^i$ can be nonzero when the weak phase difference vanishes.
Such a quantity is referred as a fake asymmetry ($CP$ conserving), which
reflects the effect of strong phases~\cite{prd84-096013,plb701-357}, instead of $CP$ violation.

\section{Numerical Analysis}\label{sec:3}

In this section, we calculate the branching rations ($\cal B$), the polarization fractions $f_{\lambda}$ and relative phases $\phi_{\parallel,\perp}$ (rad), together with TPAs, respectively.
The related input parameters for the numerical calculations are collected in Table~\ref{tab:constant}.
The decay constants are used the values from Refs.~\cite{prd76-074018,2105-03899}, while the meson masses, Wolfenstein parameters,
and the lifetimes are taken from the PDG review~\cite{pdg2020}.
We neglect uncertainties on the constants since they are negligible with respect to other sources of uncertainties.

\begin{table}
\caption{The decay constants are taken from Refs.~\cite{prd76-074018,2105-03899}.
Other parameters are from PDG 2020~\cite{pdg2020}. }
\label{tab:constant}
\centering
\begin{tabular*}{14.5cm}{@{\extracolsep{\fill}}lllll}
  \hline\hline
\text{Mass (\text{GeV})}
& $m_{B_s}=5.37$  & $m_B=5.28$  &$m_{K^{\pm}}=0.494$ \\[1ex]
\text{Wolfenstein parameters}
& $\lambda=0.22650$  & $A=0.790$  &$\bar{\rho}=0.141$ & $\bar{\eta}=0.357$ \\[1ex]
\text{Decay constants (\text{GeV})}
& $f_{B_s}=0.23$ & $f_{B}=0.21$  &$f_{\phi(1020)}=0.215$ &$f_{\phi(1020)}^T=0.186$ \\[1ex]
\text{Lifetime (ps)}
& $\tau_{B_s}=1.51$ & $\tau_{B^0}=1.52$ \\[1ex]
\hline\hline
\end{tabular*}
\end{table}

\subsection{Global fit}

According to Eqs.~(\ref{eq:phi0})-(\ref{eq:phiv}), the total amplitudes $A$ related to both longitudinal ($L$) and transverse ($N,T$) components
for the four-body decays $B^0_{(s)}\to \phi\phi \to (K^+K^-)(K^+K^-)$ can be expanded in terms of the Gegenbauer moments from the two-meson DAs.
As a result, we can decompose the squared amplitudes
into the linear combinations of the Gegenbauer moments $a^{0,T}_{2\phi}$ and their products
\begin{eqnarray}
|A^{L}|^2&=&M^L_0+a^0_{2\phi}M^L_1+(a^0_{2\phi})^2M^L_2+(a^0_{2\phi})^3M^L_3+(a^0_{2\phi})^4M^L_4,\label{eq:ampL}\\
|A^{(i)}|^2&=&M^{(i)}_0+a^T_{2\phi}M^{(i)}_1+(a^T_{2\phi})^2M^{(i)}_2+(a^T_{2\phi})^3M^{(i)}_3+(a^T_{2\phi})^4M^{(i)}_4, i=N,T.\label{eq:ampT}
\end{eqnarray}
While for three-body decays $B_{(s)}\to (\pi,K)\phi\to (\pi,K)KK$, analogously, the squared amplitudes $|A|^2$ can be paramatrised as follows
\begin{eqnarray}
|A|^2=M_0+a^0_{2\phi}M_1+(a^0_{2\phi})^2M_2.\label{eq:amp3}
\end{eqnarray}
We then compute the coefficients $M$, which involve only the Gegenbauer polynomials, to establish the database for our global fit.

Similar to the proposal in Refs.~\cite{2105-03899,2012-15074},
we adopt the standard nonlinear least-$\chi^2$ (lsq) method~\cite{Peter:2020}, in which
the $\chi^2$ function is defined for $n$ pieces of experimental data
$v_i\pm \delta v_i$ with the errors  $\delta v_i$ and the fitted corresponding theoretical
values $v^{\rm{th}}_i$ as
\begin{eqnarray} \label{eq:fit}
	\chi^2= \sum_{i=1}^{n}  \Big(\frac {v_i - v^{\rm{th}}_i}{\delta v_i}\Big)^2.
\end{eqnarray}
In general, we should include maximal amount of data in the fit in order to minimize statistical uncertainties.
However, those measurements with significance lower than 3$\sigma$ do not impose stringent constraints,
and need not be taken into account in principle.
Therefore, the Gegenbauer moments $a^{0(T)}_{2\phi}$ for the twist-2 $KK$ DAs $\phi^{0(T)}_{P}$
can be obtained by fitting the formulas in Eqs.~(\ref{eq:ampL})-(\ref{eq:amp3}) with the Gegenbauer-moment-independent database to the five pieces of
$B^{0(+)} \to K^{0(+)}\phi\to K^{0(+)} KK$ and $B^0_{s}\to \phi\phi \to (K^+K^-)(K^+K^-)$ data,
including three branching ratios and two polarization fractions as summarized in Table~\ref{tab:fitdata},
\begin{eqnarray}\label{eq:phinew}
a^0_{2\phi}=0.40\pm0.06, \quad\quad  a^T_{2\phi}=1.48\pm0.07,
\end{eqnarray}
whose errors mainly arise from experimental uncertainties.
For comparison, the updated fitting results are also listed in Table~\ref{tab:fitdata} and match well with the data within errors.

Note that our $a^0_{2\phi}$, determined with $\chi^2 / d.o.f.=1.3$,  is distinct from the value $a^0_{2\phi}=-0.31\pm0.19$ in Ref.~\cite{2105-03899},
which can be understood from the following clarification.
The additional new four-body decay $B_s^0 \to \phi\phi \to  (KK)(KK)$ included in the present work
is dominated by $B_s^0 \to (\phi \to) KK$ transition form factors,
the ${\cal B}$ of which could be more sensitive to the Gegenbauer moment $a^0_{2\phi}$.
Hence, the measured $B_s^0 \to \phi\phi \to (KK)(KK)$ branching ratio can give an effective constraint on the global fit of the $KK$ two-meson DAs,
and the corresponding  fitting result of the $a^0_{2\phi}$ could be changed a lot: from $-0.31\pm0.19$ to $0.40\pm0.06$.

One can also observe that the $a^T_{2\phi}$ fitted in this work  is slightly larger than unity as shown in Eq.~(\ref{eq:phinew}),
which is not favored in view of the convergence of the Gegenbauer expansion.
We then added one more Gegenbauer moment $a^T_{4\phi}$ in the twist-2 transverse component $\phi_P^{T}$.
Naturally, the Eq.~(\ref{eq:ampT}) should be replaced with the following form,
\begin{eqnarray}
|A^{(i)}|^2&=&M^{(i)}_0+a^T_{2\phi}M^{(i)}_1+(a^T_{2\phi})^2M^{(i)}_2
+(a^T_{2\phi})^3M^{(i)}_3+(a^T_{2\phi})^4M^{(i)}_4+(a^T_{4\phi})M^{(i)}_5+(a^T_{4\phi})^2M^{(i)}_6\nonumber\\
&+&(a^T_{4\phi})^3M^{(i)}_7+(a^T_{4\phi})^4M^{(i)}_8+(a^T_{2\phi}a^T_{4\phi})M^{(i)}_9+(a^T_{2\phi})^2(a^T_{4\phi})M^{(i)}_{10}
+(a^T_{2\phi})^3(a^T_{4\phi})M^{(i)}_{11}\nonumber\\
&+&(a^T_{2\phi})(a^T_{4\phi})^2M^{(i)}_{12}+(a^T_{2\phi})^2(a^T_{4\phi})^2M^{(i)}_{13}+(a^T_{2\phi})(a^T_{4\phi})^3M^{(i)}_{14},
\quad  i=N,T,
\end{eqnarray}
and a fit with $\chi^2 / d.o.f.=1.3$ is attained,
\begin{eqnarray}\label{eq:phia4}
a^0_{2\phi}=0.40\pm0.06, \quad\quad  a^T_{2\phi}=0.85\pm0.32,\quad\quad  a^T_{4\phi}=0.77\pm0.39.
\end{eqnarray}
The outcome of both $a^T_{2\phi}$ and $a^T_{4\phi}$ in Eq.~(\ref{eq:phia4}) are all smaller than unit,
implying that the contributions from the higher-order Gegenbauer moment is significant.
In principle, we should introduce the same number of Gegenbauer moments for the two twist-2 DAs $\phi_P^{0}$ and $\phi_P^{T}$.
However, it is not practical to include many parameters in the fit because of the limited amount of experimental data at present.
Consequently, we will adopt the $a^{0,T}_{2\phi}$ as presented in Eq.~(\ref{eq:phinew}) in this work.
Anyway, the above results show that the $a^T_{2\phi}$ can be reduced efficiently by including the higher moment $a^T_{4\phi}$ into the fit
when more experimental data with improved precision are available in the future.

\begin{table}[htbp!]
	\centering
	\caption{Experimental data for branching ratios and polarization fractions \cite{pdg2020}, and the theoretical
	results derived from the fitted Gegenbauer moments in Eq.~(\ref{eq:phinew}).
For simplicity, only the theoretical errors from the Gegenbauer moments are presented.}
    \setlength{\tabcolsep}{3mm}{
	\begin{tabular}{lccccccc}
\hline\hline
		\multirow{2}{*}{channel}         &\multicolumn{3}{c}{data}                       &\multicolumn{3}{c}{fit}\cr\cline{2-7}
	                                     &      &${\cal B}(10^{-6})$&      &   &${\cal B}(10^{-6})$&    \cr \hline
$B^{+} \to K^{+}\phi\to K^{+} KK$        &      & $8.8\pm 0.7$ &           &     & $8.8_{-0.2}^{+0.3}$    &   \\
$B^{0} \to K^{0}\phi\to K^{0} KK$        &      & $7.3\pm 0.7$ &           &     & $8.5_{-0.3}^{+0.2}$    &   \\
\hline\hline
\multirow{2}{*}{channel}                &\multicolumn{3}{c}{data}                       &\multicolumn{3}{c}{fit}\cr\cline{2-7}
	                                     &${\cal B}(10^{-6})$      & $f_0(\%)$  & $f_{\bot}(\%)$      &${\cal B}(10^{-6})$       & $f_0(\%)$  & $f_{\bot}(\%)$\cr \hline
$B_s^0 \to \phi\phi\to (KK)(KK)$          &$18.7\pm 1.5$      & $37.8\pm 1.3$ & $29.2\pm 0.9$          &$18.1^{+0.7}_{-0.8}$  & $38.2^{+2.4}_{-2.5}$ & $30.9^{+1.2}_{-0.9}$    \\

\hline\hline
\end{tabular}}
\label{tab:fitdata}
\end{table}
\subsection{$S$-wave contributions}
\begin{table}[!htbh]
\caption{PQCD predictions for the  branching ratios of various components and their
sum in the $B^0_{(s)}\rightarrow (K^+ K^-)(K^+ K^-)$ decays.
The theoretical uncertainties are attributed to the variations of the shape parameter $\omega_{B_{(s)}}$ in the $B_{(s)}$ meson DA, of the Gegenbauer moments in various twist DAs of $KK$ pair, and of the hard scale $t$ and the QCD scale $\Lambda_{\rm QCD}$.}
\label{tab:brfour}
\begin{ruledtabular}
\begin{threeparttable}
\setlength{\tabcolsep}{2.3mm}{
\begin{tabular}{lcc}
Components               & $B_s^0\rightarrow(K^+K^-)(K^+K^-)$               & $B^0\rightarrow(K^+K^-)(K^+K^-)$                          \\ \hline
$\mathcal{B}_0$          & $(1.73_{-0.43-0.13-0.65}^{+0.62+0.13+0.77})\times10^{-6}$                     &$(3.98_{-0.05-0.06-0.07}^{+0.06+0.07+0.06})\times10^{-9}$  \\
$\mathcal{B}_\parallel$  & $(1.40_{-0.09-0.10-0.59}^{+0.10+0.11+0.58})\times10^{-6}$                     &$(4.48_{-0.12-0.04-0.05}^{+0.08+0.05+0.08})\times10^{-11}$  \\
$\mathcal{B}_\perp$      & $(1.40_{-0.11-0.10-0.62}^{+0.13+0.10+0.58})\times10^{-6}$                     &$(1.01_{-0.02-0.24-0.47}^{+0.08+0.13+0.37})\times10^{-12}$  \\
$\mathcal{B}_{VS}$      &$(3.66_{-1.02-0.52-0.98}^{+1.80+0.56+1.36})\times10^{-8}$                &$(5.50_{-0.20-1.46-1.16}^{+0.60+1.70+1.30})\times10^{-11}$ \\
$\mathcal{B}_{SS}$       & $(4.38_{-1.35-1.40-2.00}^{+2.18+2.05+2.60})\times10^{-9}$               &$(1.17_{-0.20-0.59-0.20}^{+0.22+0.96+0.14})\times10^{-11}$  \\
$\mathcal{B}_{ total}$   & $(4.57_{-0.64-0.34-1.87}^{+0.86+0.35+1.95})\times10^{-6}$              &$(4.09_{-0.06-0.08-0.08}^{+0.07+0.10+0.08})\times10^{-9}$  \\
\end{tabular}}
\end{threeparttable}
\end{ruledtabular}
\end{table}

\begin{table}[!htbh]
\caption{PQCD predictions of the $S$-wave fractions in the $B^0_{(s)}\rightarrow (K^+ K^-)(K^+ K^-)$ decays.
The sources of the theoretical errors are the same as in Table~\ref{tab:brfour}.}
\label{tab:sf}
\begin{ruledtabular}
\begin{threeparttable}
\setlength{\tabcolsep}{1mm}{
\begin{tabular}{lccc}
Modes                                & $f_{VS}(\%)$            & $f_{SS}(\%)$             & $f_{S-wave}(\%)$   \\ \hline
$B_s^0\rightarrow(K^+K^-)(K^+K^-)$   & $0.801_{-0.129-0.059-0.031}^{+0.205+0.057+0.192}$ & $0.096_{-0.019-0.026-0.008}^{+0.025+0.035+0.011}$
                                     & $0.897_{-0.148-0.085-0.039}^{+0.230+0.092+0.203}$  \\
$B^0\rightarrow(K^+K^-)(K^+K^-)$     & $1.345_{-0.030-0.338-0.263}^{+0.121+0.373+0.286}$  & $0.286_{-0.045-0.141-0.044}^{+0.048+0.222+0.028}$
                                     & $1.631_{-0.075-0.479-0.307}^{+0.169+0.595+0.314}$ \\
\end{tabular}}
\end{threeparttable}
\end{ruledtabular}
\end{table}
The PQCD predictions for the branching ratios of various components and their sum in the
$B^0_{(s)}\to (K^+K^-)(K^+K^-)$ decays are summarized in Table~\ref{tab:brfour},
in which the theoretical uncertainties are estimated from three different sources.
The first error is due to the shape parameters $\omega_{B}$ in the $B_{(s)}$ meson DAs with $10\%$ variation.
The second one comes from the Gegenbauer moments in various twist DAs of $KK$ pair with different intermediate resonances.
The last one is caused by the variation of the hard scale $t$ from $0.75t$ to $1.25t$ (without changing $1/b_i$) and the QCD scale $\Lambda_{\rm QCD}=0.25\pm0.05$~GeV, which characterizes the effect of the next-to-leading-order QCD contributions.
The three uncertainties are comparable, and their combined impacts could exceed $50\%$,
implying that the nonperturbative parameters in the DAs of the initial and final states need to be constrained more precisely,
and the higher-order correction to four-body $B$ meson decays is critical.
It should be stressed that these considered modes are induced only
by penguin operators in the PQCD approach at leading order as can be seen easily from the Appendix, their direct $CP$ violations are naturally zero without the interference between the tree and penguin amplitudes.

In contrast to the vector mesons, the identification of scalar mesons is a long-standing puzzle,
and the underlying structure of scalar mesons is not theoretically well established (for a review, see Ref.~\cite{pdg2020}).
Based on the assumption that $f_0(980)$ is a pure $s\bar s$ state,
different kinds of theoretical approaches have been applied to study the $B_{(s)}$ meson decays involving $f_0(980)$ in the final states, for instance:
(a) the charmonium decay $B_s \to J/\psi f_0(980)$ was analysed  in the light-cone QCD sum rule and factorization assumption~\cite{prd81-074001},
as well as the generalized factorization and $SU(3)$ flavor symmetry~\cite{prd83-094027};
(b) in Ref.~\cite{epjc80-554}, the $\bar B_s \to f_0(980)$ transition form factor was calculated from the light-cone sum rules with $B$-meson DAs.
As a first approximation, the scalar meson $f_0(980)$ is taken into account in the $s\bar s$ density operator in the present work.
The $S$-wave time-like form factor $F_S(\omega^2)$ adopted to parameterize the $S$-wave two-kaon DAs have been determined in Refs.~\cite{prd102-056017,prd105-053002,epjc79-792}.

By using the numerical results given in Table~\ref{tab:brfour}, the $S$-wave fractions defined as
\begin{eqnarray}\label{eq:fss}
f_{\sigma}&=&\frac{\mathcal{B}_{\sigma}}{\mathcal{B}_{\text{total}}},\quad
f_{S-\text{wave}}= \sum_{\sigma}f_{\sigma},\quad \sigma={VS,SS},
\end{eqnarray}
are also calculated in our work, and the corresponding results are presented in Table~\ref{tab:sf}.
The total  $S$-wave fraction of the two considered modes is estimated to be less than $2\%$,
which is in good agreement with the experimental analysis that the contributions from the $S$-wave components are negligible~\cite{LHCb:2019jgw}.

In general, the branching ratios of two-body $B^0_{(s)} \to R_1R_2$ decays can be extracted from the corresponding four-body decay modes in Table~\ref{tab:brfour}
under the narrow width approximation
\beq
\label{2body}
\mathcal{B}(B^0_{(s)} \rightarrow R_1R_2\rightarrow (K^+K^-)(K^+K^-))&\approx& \mathcal{B}(B^0_{(s)} \rightarrow R_1R_2)\times \mathcal{B}(R_1 \rightarrow K^+K^-)\times \mathcal{B}(R_2 \rightarrow K^+K^-).
\eeq

Before evaluating ${\cal B}(f_0(980)\to K^+K^-)$, we firstly define the ratio between the $f_0(980)\to K^+K^-$ and $f_0(980)\to \pi^+\pi^-$:
\begin{eqnarray}
R_{K/\pi}=\frac{{\cal B}(f_0(980)\to K^+K^-)}{{\cal B}(f_0(980)\to \pi^+\pi^-)}.
\end{eqnarray}
In recent years, BABAR Collaboration has measured the ratio of the partial decay width of $f_0(980)\to K^+K^-$ to $f_0(980)\to \pi^+\pi^-$ of
$R^{\rm exp}_{K/\pi}=0.69\pm0.32$ using the $B\to KK^+K^-$ and $B\to K\pi^+\pi^-$ decays~\cite{prd74-032003}.
Meanwhile, BES also performed a partial wave analysis of $\chi_{c0}\to f_0(980)f_0(980)\to \pi^+\pi^-\pi^+\pi^-$
and $\chi_{c0}\to f_0(980)f_0(980)\to \pi^+\pi^-K^+K^-$ in $\psi(2S)\to \gamma\chi_{c0}$ decay and extracted the ratio as
$R^{\rm exp}_{K/\pi}=0.25^{+0.17}_{-0.11}$~\cite{prd70-092002,prd72-092002}.
Their average yielded $R^{\rm exp}_{K/\pi}=0.35^{+0.15}_{-0.14}$~\cite{prd92-032002}.
Utilizing the ${\cal B}(f_0(980)\to \pi^+\pi^-)=0.50$,
which is taken from~\cite{prd87-114001} and in agreement with the value of ${\cal B}(f_0(980)\to \pi^+\pi^-)=0.46$ reported by LHCb~\cite{prd87-052001},
one can obtain the branching fraction ${\cal B}(f_0(980)\to K^+K^-)=0.175$.

Relying on the ratio $R_{K/\pi}$ and ${\cal{B}}(\phi\to K^+K^-)=49.2\%$, the branching ratio of three-body decay $B_s^0 \to \phi (f_0(980)\to)\pi^+\pi^-$
is calculated as follows,
\begin{eqnarray}\label{eq:phipipi}
{\cal B}(B_s^0 \to \phi (f_0(980)\to)\pi^+\pi^-)=\frac{{\cal B}(B_s^0 \to \phi f_0(980)\to (K^+K^-)(K^+K^-))}{{\cal{B}}(\phi\to K^+K^-)}\cdot R_{\pi/K}
=(1.06^{+0.67}_{-0.44})\times 10^{-7}.
\end{eqnarray}
On the experimental side, the LHCb Collaboration has reported the measurement
${\cal B}(B_s^0 \to \phi(f_0(980)\to)\pi^+\pi^-)=(1.12\pm 0.21)\times 10^{-6}$,
where the region of $\pi\pi$ invariant mass is $0.4<\omega_{\pi\pi}<1.6$ ${\rm GeV}$~\cite{prd95-012006}.
It is seen that the ${\cal B}=(1.06^{+0.67}_{-0.44})\times 10^{-7}$ in Eq.~(\ref{eq:phipipi}) is almost one order of magnitude
smaller than the experimental value ${\cal B}=(1.12\pm 0.21)\times 10^{-6}$,
as well as the previous three-body PQCD result ${\cal B}=(2.35^{+4.06}_{-1.13})\times 10^{-6}$~\cite{epjc81-91}.
We have found that the contribution of scalar resonance $f_0(980)$ relies on the final-state invariant mass range strongly,
since it has a wide decay width.
For example, we recalculate the branching ratio of the four-body decay ${\cal B}(B_s^0 \to \phi f_0(980)\to (K^+K^-)(K^+K^-))=8.43\times 10^{-7}$
by enlarging the $KK$ invariant mass range of the resonance $f_0(980)$ from $[m_\phi-30{\rm MeV}, m_\phi+30 {\rm MeV}]$ to $[2m_K, m_B-m_\phi]$.
According to Eq.~(\ref{eq:phipipi}),  the relevant branching ratio of the three-body decay $B_s^0 \to \phi(f_0(980)\to)\pi^+\pi^-$
is estimated to be $4.89\times 10^{-6}$, which becomes comparable with the experimental data.

The double $S$-wave decays $B_{(s)}^0 \to f_0(980)f_0(980)$ have already been systematically studied
in the two-body framework within the PQCD approach~\cite{ epjc82-177,prd102-116007}.
Taking the  $B_s^0 \to f_0(980)f_0(980)$ decay as an example, we can roughly estimate ${\cal B}(B_s^0 \to f_0(980)f_0(980))=(1.43^{+1.29}_{-0.91})\times 10^{-7}$ from the
four-body decay $B_s^0 \to f_0(980)f_0(980)\to (K^+K^-)(K^+K^-)$ in Table~\ref{tab:brfour} on basis of  Eq.~(\ref{2body}).
It is worthwhile to note that the estimation ${\cal B}=(1.43^{+1.29}_{-0.91})\times 10^{-7}$
is much smaller than the previous two PQCD results ${\cal B}( B_s^0 \to f_0(980)f_0(980))=(2.66^{+1.08}_{-0.85})\times 10^{-4}$~\cite{prd102-116007}
and ${\cal B}(B_s^0 \to f_0(980)f_0(980))=(5.31^{+1.74}_{-1.39})\times 10^{-4}$~\cite{ epjc82-177}.
Strictly speaking, the narrow width approximation is actually not fully justified since such approximation has its scope of application.
As pointed out in Refs.~\cite{Cheng:2020iwk,Cheng:2020mna},  the narrow width approximation should be corrected by including finite-width effects
for the broad scalar intermediate state.
The two-body result extracted from the four-body branching ratio may
suffer from a large uncertainty due to the finite-width effects of the scalar resonance.
In addition, as stated above, the $S$-wave contributions show a strong dependence on the range of the $KK$ invariant mass.
We hope that the future LHCb and Belle II experiments can perform a direct measurement on four-body decays $B^0_s\to \phi f_0(980) \to (K^+K^-)(K^+K^-)$ and $B^0_s\to f_0(980) f_0(980) \to (K^+K^-)(K^+K^-)$.

\subsection{Branching ratios and polarization fractions of two-body $B^0_{(s)} \to \phi\phi$ decays}

With the narrow width approximation in Eq.~(\ref{2body}),  the branching ratios of two-body $B^0_{(s)} \to \phi\phi$ decays are extracted in Table~\ref{tab:brtwo}.
The polarization fractions of the two-body $B^0_{(s)} \to \phi\phi$ decays together with two relative phases calculated in this work are also listed in Table~\ref{tab:brtwo}.
For a comparison, we display the updated predictions in the QCDF~\cite{prd80-114026}, the previous predictions in the PQCD approach~\cite{prd91-054033,prd76-074018}, SCET~\cite{prd96-073004} and FAT~\cite{epjc77-333}.
Experimental results for branching ratios, polarization fractions and relative phases are taken from PDG 2020~\cite{pdg2020}.

\begin{table}[!htbh]
\caption{Branching ratios, polarization fractions and relative phases for the two-body $B^0_{(s)}\rightarrow \phi \phi$ decays.
For comparison, we also list the results from PQCD \cite{prd91-054033,prd76-074018},
QCDF \cite{prd80-114026}, SCET \cite{prd96-073004}, and FAT \cite{epjc77-333}.
The world averages of experimental data are taken from PDG 2020~\cite{pdg2020}.
The sources of the theoretical errors are the same as in Table~\ref{tab:brfour} but added in quadrature.}
\label{tab:brtwo}
\begin{ruledtabular}  \begin{threeparttable}
\setlength{\tabcolsep}{1mm}{ \begin{tabular}[t]{lccccc}
Modes                               & $\mathcal{B}(10^{-6})$           & $f_0(\%)$                 & $f_\perp(\%)$   & $\phi_{\|}(\rm rad)$ & $\phi_{\bot}(\rm rad)$     \\ \hline
$B_s^0\rightarrow\phi \phi$         & $18.1_{-6.1}^{+8.3}$      & $38.2_{-7.6}^{+7.2}$             &$30.9_{-2.7}^{+3.6}$ &$1.86_{-0.24}^{+0.17}$ &$1.85\pm 0.18$  \\
PQCD-I~\cite{prd76-074018}          & $35.3^{+18.7}_{-12.3}$         & $61.9_{-4.5}^{+4.4}$      & $17.4_{-1.9}^{+1.8}$ &$1.3_{-0.1}^{+0.2}$ &$1.3_{-0.1}^{+0.2}$               \\
PQCD-II \cite{prd91-054033}        & $16.7^{+4.9}_{-3.8}$             & $34.7^{+8.9}_{-7.1}$      & $31.6^{+3.5}_{-4.4}$ &$2.01\pm 0.23$ &$2.00_{-0.21}^{+0.24}$ \\
QCDF \cite{prd80-114026}                      & $16.7^{+11.6}_{-9.1}$           & $36^{+23}_{-18}$           & $\cdots$  &$\cdots$ &$\cdots$            \\
SCET \cite{prd96-073004}                        & $19.0\pm6.5$                     & $51.0\pm16.4$             & $22.2\pm9.9$  &$2.41\pm 0.62$  &$2.54\pm 0.62$      \\
FAT \cite{epjc77-333}                         & $26.4\pm7.6$                     & $39.7\pm16.0$             & $31.2\pm8.9$  &$2.53\pm 0.28$  &$2.56\pm 0.27$        \\
Data                                & $18.7\pm1.5$                     & $37.8\pm1.3$              & $29.2\pm0.9$  &$2.56\pm 0.06$  &$2.82\pm 0.19$        \\ \hline
$B^0\rightarrow\phi \phi$ & $0.016_{-0.004}^{+0.005}$ & $98.9_{-0.7}^{+0.1}$ & $0.02_{-0.00}^{+0.01}$   &$2.38_{-0.10}^{+0.21}$&$4.39_{-0.27}^{+0.21}$\\
PQCD \cite{prd91-054033}            & $0.012^{+0.006}_{-0.005}$         & $97\pm 1$ & $0.05\pm 0.02$&$3.26_{-0.14}^{+0.20}$&$3.50\pm 0.17$ \\
Data                      & $<0.027$     & $\cdots$ & $\cdots$& $\cdots$ & $\cdots$                     \\
\end{tabular}}
\end{threeparttable}
\end{ruledtabular}
\end{table}

It is obvious that most of the theoretical predictions of $B_s^0\to \phi \phi$ decay are consistent well with experiments within errors.
In Ref.~\cite{prd91-054033}, the authors kept the additional power corrections related to the ratio $r_V^2=m_V^2/m_B^2$
($m_V$ and $m_B$ denote the masses of the vector and $B$ mesons, respectively).
By including the $r_V^2$ term, their result ${\cal B}(B_s^0\to \phi\phi)=(16.7^{+4.9}_{-3.8})\times 10^{-6}$ is about twice smaller than ${\cal B}(B_s^0\to \phi\phi)=(35.3^{+18.7}_{-12.3})\times 10^{-6}$~\cite{prd76-074018},
and more close to the experimental data ${\cal B}(B_s^0\to \phi\phi)=(18.7\pm 1.5)\times 10^{-6}$~\cite{pdg2020}.
As stated in Refs.~\cite{plb763-29,prd95-056008}, the branching ratios in the quasi-two-body mechanism show their dependence on the invariant masses of the final-state meson pairs.
In this work, the factors $\eta_{1,2}=\omega_{1,2}^2/m^2_{B}$ are equal to  $r_V^2=m_V^2/m_B^2$ in Ref.~\cite{prd91-054033} when $\omega_{1,2}=m_{\phi}$.
In addition, the partonic kinematic variables have been refined to take into account finite masses of final-state mesons,
which can suppress the branching ratio of the $B_s^0\to \phi\phi$ decay effectively.
Therefore, our result agrees well with that from the updated PQCD~\cite{prd91-054033}.

The rare decay $B^0 \to \phi\phi$ can occur only via penguin annihilation topology in the SM.
The predicted branching ratio is very small at ($10^{-8}$), which makes it sensitive to any new physics contributions.
The current experiment gives the upper limit: $\mathcal{B}(B^0 \to \phi\phi) < 2.7\times10^{-8}$ at $90\%$ CL~\cite{LHCb:2019jgw}, so the more accurate
experimental results are needed to test the theory.
It is observed that the branching fraction of $B^0 \to \phi\phi$ decay is much smaller than that of $B_s^0 \to \phi\phi$ decay by almost three orders.
There are two main reasons:
the one is that the $B^0 \to \phi\phi$ governed by $b\to d$ transition is highly suppressed by the CKM matrix elements $|V_{td} /V_{ts}|^2\sim 0.05$,
the other is that $B^0 \to \phi\phi$ belongs to the pure annihilation decay.
As is known, the contributions from the annihilation diagrams (Figs.~\ref{fig2}(e)-~\ref{fig2}(h)) are always power suppressed compared to the factorizable
emission diagrams (Figs.~\ref{fig2}(a) and \ref{fig2}(b)) in the PQCD approach .
What's more, there exists a big cancellation between the two factorizable annihilation diagrams Figs.~\ref{fig2}(e) and \ref{fig2}(f) for the contributions from the $(V-A)(V-A)$ operators,
especially when two final state mesons are identical, like $B^0 \to \phi \phi$.

For the charmless $B$ decays, it is naively expected that the helicity amplitudes $H_{i}$ (with helicity $i=0,-,+$) satisfy the hierarchy pattern
\begin{align}
H_0:H_-:H_+=1: \frac{\Lambda_{QCD}}{m_b}:(\frac{\Lambda_{QCD}}{m_b})^2,
\label{eq:hierarchy}
\end{align}
which are related to the spin amplitudes $(A_{0}, A_{\parallel}, A_{\perp})$ in Appendix by
\beq
A_0=H_0, \quad  A_{\parallel}=\frac{H_{+}+H_{-}}{\sqrt{2}},\quad A_{\perp}=\frac{H_{+}-H_{-}}{\sqrt{2}}.
\eeq
The above hierarchy relation satisfies the expectation in the factorization assumption that the longitudinal
polarization should dominate based on the quark helicity analysis~\cite{zpc1-269,prd64-117503}.
In sharp contrast to these expectations, roughly equal longitudinal and transverse components are found in measurements of $B\to K^*\phi$, $B\to K^*\rho$ decays~\cite{prl91-201801,prd78-092008,prd85-072005,LHCb:2014xzf}.
Measurements of the low longitudinal polarization fraction in $B_s^0 \to \phi \phi$ by CDF~\cite{prl107-261802} and LHCb~\cite{plb713-369,prd90-052011,LHCb:2019jgw} indicate a large transverse polarization.
The longitudinal polarization fraction $f_0=0.381\pm0.007\pm0.012$ of the $B^0_s \to \phi\phi$ decay has been reported by LHCb recently~\cite{LHCb:2019jgw}, where the first uncertainty is statistical and the second systematic.
This shows that the scaling behavior shown in Eq.~(\ref{eq:hierarchy}) is violated.
The interest in the polarization in penguin transition, such as $b\to s$ decay $B_s^0 \to \phi\phi$, is motivated by its potential sensitivity to physics beyond the SM.

As shown in Table~\ref{tab:brtwo}, we obtain the longitudinal polarization fraction $f_0=(38.2^{+7.2}_{-7.6})\%$ of the $B^0_s \to \phi\phi$ decay with the updated Gegenbauer moments of two-meson DAs, which is consistent with the previous PQCD calculation~\cite{prd91-054033} and those from QCDF~\cite{prd80-114026}, SCET~\cite{prd96-073004} and FAT~\cite{epjc77-333} within uncertainties.
In the PQCD approach, the large transverse polarization fraction can be interpreted on the basis of the chirally enhanced annihilation diagrams, especially the $(S-P)(S+P)$ penguin annihilation, introduced by the QCD penguin operator $O_6$~\cite{prd71-054025}, which is originally introduced in Ref.~\cite{plb601-151}.
A special feature of the $(S-P)(S+P)$ penguin annihilation operator is that the light quarks in the final states are not produced through chiral currents. So, there is no suppression to the transverse polarization caused by the helicity flip. Then the polarization fractions satisfy $f_0\approx f_T$, with $f_T=f_\|+f_\bot$.

For the $B^0 \to \phi\phi$ decay, the longitudinal polarization contribution is dominant,
which is consistent with the recent updated PQCD calculation~\cite{prd91-054033} and also verified in Ref.~\cite{prd89-014003}.
As clarified before,  the contributions from the factorizable annihilation diagrams (Figs.~\ref{fig2}(e)-~\ref{fig2}(f)) are canceled
by each other because of the current conservation.
Hence, the terms like $F^{LL,0(\parallel)}_{a\phi}$ and $F^{LR,0(\parallel)}_{a\phi}$ in the Eq.~(\ref{eq:b0phiphi}) are exactly equal to zero,
while the only left parts $F^{LL,\perp}_{a\phi}$ and $F^{LR,\perp}_{a\phi}$ for the factorizable emission diagrams are power suppressed.
For the non-factorizable annihilation diagrams (Figs.~\ref{fig2}(g)-~\ref{fig2}(h)), the longitudinal parts give the leading and dominant contributions,
and other terms related to parallel ($M^{LL,\|}_{a\phi}$, $M^{SP,\|}_{a\phi}$) and perpendicular ($M^{LL,\bot}_{a\phi}$, $M^{SP,\bot}_{a\phi}$) components are all power suppressed.
Thus, the total transverse contributions are actually negligible, leading to $f_0 \sim 1$ as shown in Table~\ref{tab:brtwo}.

The relative phases $\phi_{\parallel}$ and $\phi_{\perp}$ of the $B^0_{(s)}\to \phi\phi$ decays are also studied in the present work as shown in Table~\ref{tab:brtwo}.
In fact, two relative phases derived from the decay amplitudes $A_{0,\parallel,\perp}$ in Eq.~(\ref{pol}) are dependent on the invariant mass $\omega_{1,2}$.
We fix $\omega_{1,2}=m_\phi$ in our calculation for comparison with the two-body analysis.
For $B^0_{s}\to \phi\phi$ decay, the differences between the two previous PQCD results~\cite{prd91-054033,prd76-074018} are mainly attributed to the
treatment of the terms in the decay amplitude proportional to the ratio $r_V^2=m_V^2/m_B^2$, which has been neglected in Ref.~\cite{prd76-074018}.
While in our calculations, the factors $\eta_{1,2}=\omega_{1,2}^2/m^2_{B}$ given in Eq.~(\ref{eq:epsilon}) become equal to $r_V^2=m_V^2/m_B^2$ in Ref.~\cite{prd91-054033} when $\omega_{1,2}=m_{\phi}$.
Therefore, our new four-body computation $\phi_{\parallel}=(1.86^{+0.17}_{-0.24} )$ ${\rm rad}$ and $\phi_{\perp}=(1.85\pm 0.18)$ $ {\rm rad}$
agree with the updated PQCD results $\phi_{\parallel}=(2.01\pm 0.23)$ $ {\rm rad}$ and $\phi_{\perp}=(2.00_{-0.21}^{+0.24})$ $ {\rm rad}$ ~\cite{prd91-054033}
within errors.
It is obvious that our predictions of the relative phases are smaller than those of the SCET~\cite{prd96-073004} and FAT~\cite{epjc77-333} calculations as well as the experimental data~\cite{pdg2020}.
As stressed above, we use the two decay
constants $f_{\phi}^{(T)}$ of the intermediate particle to determine the ratio $F_P^{\perp}(\omega^2)/F_P^{\parallel}(\omega^2)\approx (f_{\phi}^T/f_{\phi})$.
To be honest, we have omitted the phase difference between the two form factors $F_P^{\perp}(\omega^2)$ and $F_P^{\parallel}(\omega^2)$ due to the limited studies on the form factor $F_P^{\perp}(\omega^2)$.
We have found that the gap between our predictions and the measurements of two relative phases can be resolved effectively by introducing an additional phase $\beta$ in the above approximate equation,
\begin{eqnarray}
\frac{F_P^{\perp}(\omega^2)}{F_P^{\parallel}(\omega^2)}\approx \frac{f_{\phi}^T}{f_{\phi}}e^{i\beta}.
\end{eqnarray}
In Table~\ref{tab:phase}, we have examined the dependencies of two relative phases $\phi_{\parallel,\perp}$ on $\beta$ ( $\beta \in [1.0,1.6]$ ${\rm rad}$ )
and found that the $\phi_{\parallel}=2.55$ $ {\rm rad}$ and $\phi_{\perp}=2.46$ $ {\rm rad}$, obtained for $\beta=1.3$ ${\rm rad}$,  are well matched to the
data $\phi_{\parallel}=(2.56\pm 0.06)$ $ {\rm rad}$ and $\phi_{\perp}=(2.82\pm 0.19)$ $ {\rm rad}$ within  errors.
Nonetheless, it is not appropriate for us to include the parameter $\beta$ in the present fit due to the limited data.
For the pure annihilation decay $B^0 \to \phi\phi$, however,
the new four-body predictions $\phi_{\perp}=(4.39^{+0.21}_{-0.27})$ $ {\rm rad}$ and $\phi_{\parallel}=(2.38^{+0.21}_{-0.10})$ $ {\rm rad}$
are quite different from those of two-body results $\phi_{\parallel}=(3.26_{-0.14}^{+0.20})$ ${\rm rad} $ and
$\phi_{\perp}=(3.50\pm 0.17)$ ${\rm rad}$~\cite{prd91-054033}.
The main reason is that we have kept track of the additional higher power corrections related to the momenta fraction $x_B$,
which has been ignored in Ref.~\cite{prd91-054033}.
We have reexamined the two phases $\phi_{\parallel,\perp}$ without the contributions from the $x_B$:
$\phi_{\parallel}=2.67$ ${\rm rad} $, $\phi_{\perp}=2.75$ ${\rm rad}$, which are similar to the two body analysis.
It should be stressed that the contributions from the annihilation diagrams (Figs.~\ref{fig2}(e)-~\ref{fig2}(h)) are of higher power themselves
for a pure annihilation decay mode without chiral enhancement, like $B^0 \to \phi\phi$.
In that case, the terms proportional to $x_B$ in the amplitudes are not negligible and should be reserved in the calculations.
Anyway, all these theoretical predictions need to be further tested in the future when more data are available.

\begin{table}[!htbh]
\caption{The dependencies of the  $\phi_{\parallel}$ $ ({\rm rad})$ and $\phi_{\perp}$ $ ({\rm rad})$ on $\beta$ $ ({\rm rad})$ at $\omega_{1,2}=m_\phi$ in the $B^0_s\to \phi\phi$ decay, where $\beta\in [1.0,1.6] $ ${\rm rad}$.}
\label{tab:phase}
\begin{ruledtabular}
\begin{threeparttable}
\setlength{\tabcolsep}{1mm}{
\begin{tabular}{lccccccc}
$\beta$                          & 1.0   &1.1    &1.2   &1.3  &1.4   &1.5  &1.6   \\ \hline
$\phi_{\parallel}$               & $2.07$ & $2.18$ & $2.34$& $2.55$& $2.90$& $3.59$ & $4.36$ \\
$\phi_{\perp}$                   & $2.03$ & $2.13$ & $2.26$& $2.46$& $2.82$& $3.49$ &$4.35$ \\
\end{tabular}}
\end{threeparttable}
\end{ruledtabular}
\end{table}

\subsection{Triple  product asymmetries in $B^0_{(s)} \to (K^+K^-)(K^+K^-)$ decays}

In the involved neutral intermediate states $B^0_s \to\phi\phi$ and $B^0 \to \phi\phi$ modes,  each helicity amplitude involves the same single weak phase in the SM.
This results in $\mathcal{A}_\text{T-true}^i=0$  due to the vanishing weak phase difference.
The ``true" TPAs for these neutral modes are thus predicted to be zero in the SM as shown in Table~\ref{tab:tpas}.
If such asymmetries are observed experimentally, it is probably signify the presence of new physics.
On the experimental side, the measurements of TPAs for $B_s^0 \to \phi\phi \to (K^+ K^-)(K^+K^-)$ have been reported by CDF~\cite{prl107-261802} and LHCb Collaborations~\cite{plb713-369,prd90-052011} and have shown no evidence of deviations from the SM.
The most recent measurements of the ``true" TPAs give~\cite{LHCb:2019jgw}
\beq
\mathcal{A}_V&=&-0.014\pm0.011(stat)\pm0.004(syst),  \non
\mathcal{A}_U&=&-0.003\pm0.011(stat)\pm0.004(syst),
\eeq
where the first uncertainty is statistical and the second systematic.
No evidence for $CP$ violation is found, which is consistent with SM predictions.

The predicted ``fake" TPAs for the  $B^0_{(s)} \to (K^+K^-)(K^+K^-)$  decays are presented in Table~\ref{tab:tpas}.
As ``fake'' TPAs are due to strong phases and require no $CP$ violation, the large fake $\mathcal{A}_{\text{T-fake}}^{1,2}$ simply reflects the importance of the strong final-state phases.
The magnitude of $\mathcal{A}_\text{T-fake}^1$ for the $B^0_s$ channel exceeds ten percent and reaches $30.4\%$.
The sizable magnitude is mainly enhanced by the strong phase difference between the longitudinal and perpendicular polarization amplitudes, which is found in Table~\ref{tab:brtwo}.
The smallness of $\mathcal{A}_\text{T-fake}^2$ is attributed to the suppression from the strong phase
difference  between the perpendicular and parallel polarization amplitudes, which can be seen in Table~\ref{tab:brtwo} and has been verified by LHCb~\cite{LHCb:2019jgw}.
Hence, observations of $\mathcal{A}_\text{T-fake}^2$ with large values would signal new physics beyond the SM.
As mentioned above, the hierarchy in Eq.~(\ref{eq:hierarchy}) is numerically not respected by penguin-dominated decays.
Thus, final states with large transverse amplitude fractions are favourable for the measurement of TPAs and can provide valuable complementary information on $CP$ violation without requiring the generation of a sizable strong phase difference.
Our predictions can be tested in the future.

\begin{table}[htbp!]
	\centering
	\caption{PQCD predictions for the TPAs ($\%$) of the four-body $B^0_{(s)}\rightarrow (K^+ K^-)(K^+ K^-)$ decays.
The sources of theoretical errors are same as in Table~\ref{tab:brfour} but added in quadrature. }
    \setlength{\tabcolsep}{2.3mm}{
	\begin{tabular}{lccccc}
\hline\hline
\multirow{2}{*}{channel}                &\multicolumn{4}{c}{$\text{TPAs}$-1}               \cr\cline{2-5}
	                 &$\mathcal{A}_{\text{T}}^1$&$\bar{\mathcal{A}}_{\text{T}}^1$&$\mathcal{A}_{\text{T-true}}^1$&$\mathcal{A}_{\text{T-fake}}^1$ \cr \hline
$B^0_s \to \phi\phi \to (K^+K^-)(K^+K^-)$ &$30.38^{+1.16}_{-2.39}$&$-30.38^{+2.39}_{-1.16}$&$0$&$30.38^{+1.16}_{-2.39}$\\
$B^0 \to \phi\phi \to (K^+K^-)(K^+K^-)$ &$0.67^{+0.21}_{-0.14}$&$-0.67_{-0.21}^{+0.14}$&$0$&$0.67^{+0.21}_{-0.14}$  \\
\hline\hline
\multirow{2}{*}{channel}                &\multicolumn{4}{c}{$\text{TPAs}$-2}                \cr\cline{2-5}
	                 &$\mathcal{A}_{\text{T}}^2$&$\bar{\mathcal{A}}_{\text{T}}^2$&$\mathcal{A}_{\text{T-true}}^2$&$\mathcal{A}_{\text{T-fake}}^2$\cr \hline
$B^0_s \to \phi\phi \to (K^+K^-)(K^+K^-)$ &$0.15^{+0.03}_{-0.10}$&$-0.15_{-0.03}^{+0.10}$&$0$&$0.15^{+0.03}_{-0.10}$   \\
$B^0 \to \phi\phi \to (K^+K^-)(K^+K^-)$  &$-0.11^{+0.02}_{-0.04}$&$0.11_{-0.02}^{+0.04}$&$0$&$-0.11^{+0.02}_{-0.04}$    \\

\hline\hline
\end{tabular}}
\label{tab:tpas}
\end{table}

\section{Conclusion}\label{sec:4}
In this work, we have studied the related helicity amplitudes of four-body $B^0_{(s)} \to (K^+K^-)(K^+K^-)$ decays based on the angular analysis, where $K^+K^-$ invariant-mass spectrum is dominated by the vector resonance $\phi$.
The scalar resonance $f_0(980)$ is also contributed in the $K^+K^-$ invariant-mass range.
The strong dynamics of the scalar or vector resonance decays into the meson pair
is parametrized into the corresponding two-meson distribution amplitude,
which has been established in three-body $B$ meson decays and further improved by performing a global fit through combining the measured branching ratios in four-body decays.

The branching ratios of four-body $B^0_{(s)} \to (K^+K^-)(K^+K^-)$ decays are presented with the updated $P$-wave two-kaon distribution amplitudes.
We have extracted the two-body $B_{(s)}^0 \rightarrow \phi\phi$ branching ratios from the results for the
corresponding four-body decays under the narrow-width approximation and shown the polarization fractions and relative phases of the decay channels.
The obtained two-body branching ratios agree well with previous theoretical studies in the
two-body framework within errors.
The predicted hierarchy pattern for the longitudinal polarization fractions in the $B^0_{(s)} \to \phi\phi$ decays is in agreement with the data.

Since the triple product asymmetries are helpful to discover physics beyond the standard model, we perform an angular analysis and estimate the triple product asymmetries on four-body $B_{(s)}^0 \to \phi\phi \to (K^+K^-)(K^+K^-)$ decays.
The ``true" TPAs of four-body $B^0_s \to \phi\phi \to (K^+K^-)(K^+K^-)$ decays are predicted to be zero due to the vanishing weak phase difference, which is consistent with the experiments.
The prediction of ``fake" TPA $\mathcal{A}_\text{T-fake}^1$ of $B^0_s \to \phi\phi \to (K^+K^-)(K^+K^-)$ reaches $30\%$ in magnitude, which reflects the importance of the strong final-state phases and can be tested in the future.
We also make predictions of TPAs for $B^0 \to \phi\phi \to (K^+K^-)(K^+K^-)$ decay and wait for the confrontation with future data.

\begin{acknowledgments}
Many thanks to H.n.~Li for valuable discussions.
This work was supported by the National Natural Science Foundation of China under the No.~12005103, No.~12075086, No.~11775117, No.~12105028.
YL is also supported by the Natural Science Foundation of Jiangsu Province under Grant No.~BK20190508 and the Research Start-up Funding of Nanjing Agricultural University.
DCY is also supported by the Natural Science Foundation of Jiangsu Province under Grant No.~BK20200980.
ZR is supported in part by the Natural Science Foundation of Hebei Province under Grant No.~A2019209449 and No.~A2021209002.

\end{acknowledgments}

\appendix
\section{Decay amplitudes}
In the Appendix, we present the PQCD factorization formulas for the amplitudes of
the considered four-body hadronic $B$ meson decays:
\begin{itemize}
\item[]
$\bullet$ $ B \to \phi\phi\to(K^+K^-)(K^+K^-)$ decay modes ($h=0,\|,\perp$)
\begin{eqnarray}
\label{eq:b0phiphi}
\sqrt{2}A_h(B^0 \to \phi\phi\to (K^+K^-) (K^+K^-))&=& -\frac{2G_F} {\sqrt{2}}V_{tb}^*V_{td}\Big [ \left  (C_3+\frac{C_4}{3}-\frac{C_9}{2}-\frac{C_{10}}{6} \right )F^{LL,h}_{a\phi}  \non
 &+&  \left( C_4-\frac{C_{10}}{2} \right )M^{LL,h}_{a\phi}+\left (C_6-\frac{C_{8}}{2} \right ) M^{SP,h}_{a\phi}\non
 &+&\left(C_5+\frac{C_6}{3}-\frac{C_7}{2}-\frac{C_{8}}{6} \right ) F^{LR,h}_{a\phi} \Big ] ,
\end{eqnarray}

\begin{eqnarray}
\sqrt{2}A_h(B^0_s \to \phi\phi\to (K^+K^-) (K^+K^-))&=&-\frac{2G_F} {\sqrt{2}}V_{tb}^*V_{ts}\Big[ \left (C_5-\frac{C_7}{2}\right)
\left(M^{LR,h}_{e\phi}+M^{LR,h}_{a\phi}\right )   \non
&+&\frac{4}{3}\left (C_3+C_4-\frac{C_9}{2}-\frac{C_{10}}{2}\right )\left (F^{LL,h}_{e\phi}+F^{LL,h}_{a\phi}\right )\non
&+&\left ( C_3+C_4-\frac{C_9}{2}-\frac{C_{10}}{2} \right )\left (M^{LL,h}_{e\phi}+M^{LL,h}_{a\phi}\right )\non
&+& \left (C_5+\frac{C_6}{3}-\frac{C_7}{2}-\frac{C_{8}}{6}\right )\left (F^{LR,h}_{e\phi}+F^{LR,h}_{a\phi} \right )\non
&+& \left (C_6-\frac{C_8}{2}\right )\left (M^{SP,h}_{e\phi}+M^{SP,h}_{a\phi}\right ) \nonumber\\
&+&\left (C_6+\frac{C_5}{3}-\frac{C_8}{2}-\frac{C_{7}}{6} \right )F^{SP,h}_{a\phi}\Big ],
\end{eqnarray}

\item[]
$\bullet$ $ B \to f_0(980) \phi\to(K^+K^-)(K^+K^-)$ decay modes
\begin{eqnarray}
A(B^0 \to f_0\phi\to (K^+K^-) (K^+K^-))&=& -\frac{G_F} {\sqrt{2}}V_{tb}^*V_{td}\Big [\left (C_4-\frac{C_{10}}{2}\right )\left (M^{LL}_{a\phi}+M^{LL}_{af_0}\right ) \non
&+&\left (C_3+\frac{C_4}{3}-\frac{C_9}{2}-\frac{C_{10}}{6}\right )\left (F^{LL}_{a\phi}+F^{LL}_{af_0}\right )\non
&+&\left (C_5+\frac{C_6}{3}-\frac{C_7}{2}-\frac{C_{8}}{6}\right )\left (F^{LR}_{a\phi}+F^{LR}_{af_0}\right )\nonumber\\
&+&\left (C_6-\frac{C_{8}}{2}\right )\left (M^{SP}_{a\phi}+M^{SP}_{af_0}\right ) \Big] ,
\end{eqnarray}
\begin{eqnarray}
A(B_s^0 \to f_0\phi \to(K^+K^-) (K^+K^-))&=&-\frac{2G_F} {\sqrt{2}}V_{tb}^*V_{ts}\Big [\left (C_5-\frac{C_7}{2}\right )\left (M^{LR}_{e\phi}+M^{LR}_{a\phi}+M^{LR}_{ef_0}+M^{LR}_{af_0}\right)\non
&+&\frac{4}{3} \left (C_3+C_4-\frac{C_9}{2}-\frac{C_{10}}{2} \right ) \left (F^{LL}_{ef_0}+F^{LL}_{a\phi}+F^{LL}_{af_0}\right )\non
&+& \left(C_6-\frac{C_8}{2}\right)\left(M^{SP}_{ef_0}+M^{SP}_{af_0}+M^{SP}_{e\phi}+M^{SP}_{a\phi}\right)\non
&+& \left ( C_3+C_4-\frac{C_9}{2}-\frac{C_{10}}{2} \right ) \left (M^{LL}_{e\phi}+M^{LL}_{a\phi}+M^{LL}_{ef_0}+M^{LL}_{af_0} \right )\non
&+&\left (C_5+\frac{C_6}{3}-\frac{C_7}{2}-\frac{C_{8}}{6}\right ) \left (F^{LR}_{ef_0}+F^{LR}_{a\phi}+F^{LR}_{af_0}\right )\non
&+&\left ( C_6+\frac{C_5}{3}-\frac{C_8}{2}-\frac{C_{7}}{6}\right)\left (F^{SP}_{e\phi}+F^{SP}_{a\phi}+F^{SP}_{af_0}\right )\Big ],
\end{eqnarray}


\item[]
$\bullet$ $B \to f_0(980) f_0(980)\to(K^+K^-)(K^+K^-)$ decay modes
\begin{eqnarray}
\sqrt{2}A(B^0 \to f_0f_0\to (K^+K^-) (K^+K^-))&=& -\frac{2G_F} {\sqrt{2}}V_{tb}^*V_{td} \Big [\left (C_3+\frac{C_4}{3}-\frac{C_9}{2}-\frac{C_{10}}{6}\right )F^{LL}_{af_0}\non
 &+&\left (C_5+\frac{C_6}{3}-\frac{C_7}{2}-\frac{C_{8}}{6}\right )F^{LR}_{af_0}\non
 &+&\left (C_4-\frac{C_{10}}{2}\right )M^{LL}_{af_0}+\left (C_6-\frac{C_{8}}{2}\right )M^{SP}_{af_0} \Big], \qquad
\end{eqnarray}

\begin{eqnarray}
\sqrt{2}A(B_s^0 \to  f_0f_0\to (K^+K^-) (K^+K^-))&=&-\frac{2G_F} {\sqrt{2}}V_{tb}^*V_{ts}\Big[\frac{4}{3}\left (C_3+C_4-\frac{C_9}{2}-\frac{C_{10}}{2}\right )F^{LL}_{af_0}\non
&+&\left (C_5+\frac{C_6}{3}-\frac{C_7}{2}-\frac{C_{8}}{6}\right )F^{LR}_{af_0}\non
&+& \left(C_6-\frac{C_8}{2}\right )\left (M^{SP}_{ef_0}+M^{SP}_{af_0}\right ) \non
&+&\left (C_5-\frac{C_7}{2}\right )\left (M^{LR}_{ef_0}+M^{LR}_{af_0}\right)\non
 &+&\left (C_6+\frac{C_5}{3}-\frac{C_8}{2}-\frac{C_{7}}{6}\right )\left (F^{SP}_{ef_0}+F^{SP}_{af_0}\right)\non
 &+& \left (C_3+C_4-\frac{C_9}{2}-\frac{C_{10}}{2}\right )\left (M^{LL}_{ef_0}+M^{LL}_{af_0}\right )
 )\Big],\non
\end{eqnarray}

\end{itemize}
where $G_F=1.16639\times 10^{-5}$ GeV$^{-2}$ is the Fermi coupling constant and the $V_{ij}$'s are the Cabibbo-Kobayashi-Maskawa matrix elements.
The superscripts $LL$, $LR$, and $SP$ refer to the contributions from $(V-A)\otimes(V-A)$, $(V-A)\otimes(V+A)$, and $(S-P)\otimes(S+P)$ operators, respectively.
The explicit formulas for the factorizable emission (annihilation) contributions $F_{e(a)}$ and the nonfactorizable emission (annihilation) contributions
$M_{e(a)}$ from Fig.~\ref{fig2} can be obtained easily in Ref.~\cite{zjhep}.


\end{document}